\documentclass{optica-article}

\journal{opticajournal} 

\articletype{Research Article}

\usepackage{lineno}
\usepackage{orcidlink}
\usepackage{siunitx}
\usepackage{comment}
\usepackage{titling}

\usepackage{bookmark}
\bookmarksetup{
 numbered,
 open
}
\renewcommand*{\thesection}{\arabic{section}}

\begin{document}

\title{Spectral evolution of two-photon emission in microresonators}

\author{Francesca Famà\,\orcidlink{0009-0009-3162-8806},\authormark{1, *} Stefano Dello Russo,\orcidlink{0000-0002-0034-4749}, \authormark{2} Stefano Giaccari,\orcidlink{0000-0001-9467-2403},\authormark{1} Salvatore Virzì,\orcidlink{0000-0002-9067-5970},\authormark{1} Lorenzo Lucia,\orcidlink{0009-0004-6534-715X},\authormark{1} Cecilia Clivati,\orcidlink{0000-0002-7289-6403},\authormark{1} Chiara Gionco,\orcidlink{0000-0002-5351-8116},\authormark{1} Mario Siciliani de Cumis,\orcidlink{0000-0003-2854-5881},\authormark{2} Stefano Condio,\orcidlink{0000-0003-4921-7959},\authormark{1} Erik Cerrato,\orcidlink{0000-0002-6337-8751},\authormark{1} Gianluca Bertaina,\orcidlink{0000-0002-9440-4537},\authormark{1} Alice Meda,\orcidlink{0000-0001-5658-3194},\authormark{1} Marco Gramegna,\orcidlink{0000-0002-5725-0444},\authormark{1} Marco Genovese,\orcidlink{0000-0001-9186-8849},\authormark{1} Filippo Levi,\orcidlink{0000-0002-0206-9082},\authormark{1} Ivo Pietro Degiovanni,\orcidlink{0000-0003-0332-3115},\authormark{1} and Davide Calonico\orcidlink{0000-0002-0345-859X},\authormark{1}}

\address{\authormark{1}Istituto Nazionale di Ricerca Metrologica, Strada delle Cacce 91, 10135 Turin, Italy\\
\authormark{2}Agenzia Spaziale Italiana, Centro di Geodesia Spaziale “G. Colombo”, 75100 Matera, Italy\\}

\email{\authormark{*}f.fama@inrim.it} 

\begin{abstract*} 
High-Q silicon nitride microresonators are versatile sources for generating photon pairs via four-wave mixing. We investigate the spectral coherence of this process, tracking the transition from the spontaneous quantum regime to the onset of optical parametric oscillation. By combining time-correlation measurements with phase-sensitive measurements, we continuously monitor the emission linewidth as it evolves from a cavity-lifetime-limited linewidth toward the pump-linewidth scale. This characterization is essential for optimizing integrated sources for scalable quantum networks.
\end{abstract*}

\setcounter{section}{0}
\renewcommand*{\theHsection}{chX.\the\value{section}}

\section{\label{sec:level1}Introduction}

Two-photon sources have played a central role in the development of modern quantum science, providing versatile and reliable platforms for the generation of entangled photon pairs and heralded single-photon states~\cite{Cout23, Rari86, Brid11}. 

 Since their first experimental demonstrations, they have enabled controlled access to nonclassical properties of light and have become indispensable tools for fundamental studies, including landmark tests of Bell’s inequalities~\cite{Asp82, Giu13, Sha15, Wen19, Genovese1, Vir24} that revealed the nonlocal nature of quantum correlations. At the same time, they paved the way for protocols such as quantum teleportation~\cite{Mar03, Yin12}, entanglement swapping~\cite{Jen02, Tak02}, and dense coding~\cite{Matt96}. Two-photon sources also underpin a broad range of applications, from entanglement-based quantum key distribution (QKD), including device-independent~\cite{Thew07,Zapa23} and measurement-device-independent protocols~\cite{Yan25}, to sub-shot-noise measurements~\cite{Okam13, Meda17, Genovese2}, quantum-enhanced metrology and sensing beyond classical limits~\cite{Poly09,Obri09,Flam19, Genovese3}.

Experimentally, two-photon sources are commonly realized using nonlinear optical processes such as spontaneous parametric down-conversion in second-order media and four-wave mixing (FWM) in third-order platforms~\cite{Boyd08}.

While bulk nonlinear crystals have served as the foundation for these sources, integrated photonic platforms have recently garnered significant attention due to their scalability and versatility~\cite{wal15, Che16}. 
Particularly interesting are CMOS-compatible architectures, which offer the prospect of scalable and cost-effective fabrication through established semiconductor manufacturing processes. In this context, silicon nitride ($\text{Si}_3\text{N}_4$)  microresonators have emerged as a powerful platform, owing to their low optical losses, broad transparency window spanning the telecom spectrum, and compatibility with integrated photonic technologies~\cite{Levy2010}.

Owing to its centrosymmetric structure, $\text{Si}_3\text{N}_4$ lacks second-order nonlinearity~\cite{Zhang23}, making third-order FWM the primary nonlinear mechanism~\cite{Ikeda08}. In this process, two pump photons are annihilated to generate an equidistant signal-idler photon pair. When the generated frequencies coincide with resonator modes, the process is strongly enhanced, producing multiple sidebands equally spaced around the pump wavelength. This mechanism is widely used to generate photon pairs in the telecom C-band, with the signal and idler modes designed to match the standard \qty{100}{GHz} WDM grid by engineering the resonator free spectral range through the choice of resonator radius~\cite{Kras19, Ong26}. 

In the regime of low pump powers, these devices serve as sources of entangled photons and heralded single-photon states~\cite{Ma18, Cas17, Mah23} while, at high pump powers, they can enter the optical parametric oscillation (OPO) regime, where intensity correlations between the oscillating modes enable the generation of macroscopic squeezed states~\cite{Bra21}. Under suitable dispersion and detuning conditions, the same nonlinear dynamics can also support the formation of coherent Kerr frequency combs~\cite{Kipp18}, which have emerged as a key ingredient for compact optical clocks and precision metrology. 

In both the classical and quantum regimes, the spectral properties of the emitted light play a central role. They are among the factors determining the coherence of frequency combs at high power~\cite{Erk:14}, while, at the single-photon level, they set the indistinguishability and phase stability of entangled photons. Precise control over these spectral features is therefore crucial, not only for applications in classical metrology, but also for scalable quantum communication, where narrowband, phase-stable emission is particularly critical for long-distance, phase-encoded QKD or efficient coupling to quantum memories~\cite{Fek2013, Cli22}. 
However, a detailed experimental mapping of the spectral evolution across these operating regimes is currently lacking.

\begin{figure}[t]
\includegraphics[width=1\linewidth, keepaspectratio]{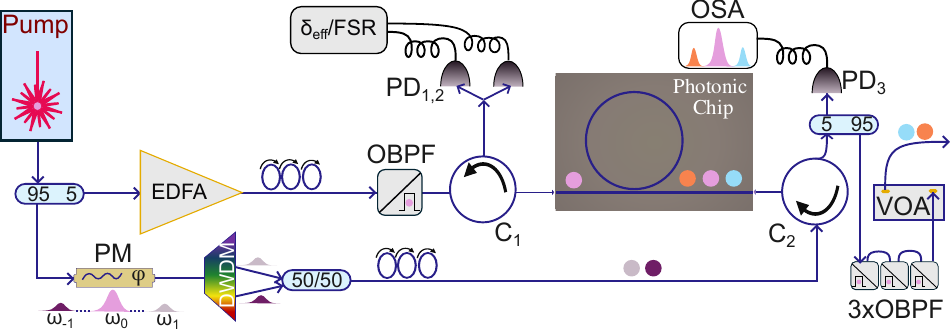}
\caption{\label{fig:Sec_exp} Schematic of the experimental setup for photon-pair generation and effective detuning measurement. A continuous-wave pump laser is split: 95$\%$ is phase-modulated to create weak fourth-order sidebands for cavity monitoring, while the remaining 5$\%$ is amplified and filtered to drive photon-pair generation in the microresonator. Two optical circulators are used to route the counterpropagating beams through the microring and detection paths. DWDM Dense wavelength division multiplexing, OBPF Optical Band Pass Filter, PD photodiode, C circulator, PM phase modulator, VOA variable optical attenuator.}
\end{figure}

In this work, we investigate both theoretically and experimentally the spectral and coherence properties of photon pairs generated via FWM in a silicon nitride microresonator driven by a continuous-wave (CW) pump laser. By adjusting the effective detuning between the pump frequency and the thermally shifted cavity resonance, we control the circulating power in the resonator and continuously transition from spontaneous FWM to the onset of OPO. 
In the former, the emission linewidth is determined by the cavity lifetime while in the latter, we observe a spectral narrowing that approaches the pump laser linewidth.

The spectral properties of the generated light across these regimes are measured by combining time-correlation measurements (cross- and auto-correlation) with phase-sensitive techniques (heterodyne beating and self-delayed interferometry).
The former probe second-order coherence functions by analyzing temporal correlations between photon detection events and provide indirect spectral information, particularly in the quantum regime where signal and idler photons are intrinsically correlated. By contrast, the latter directly probe first-order coherence and remain effective across both quantum and classical regimes, with heterodyne ultimately limited only by signal intensity. Together, these methods provide a comprehensive characterization of coherence across the classical-to-quantum transition, enabling a deeper understanding of integrated photon-pair sources.

Focusing on the below-threshold regime, where a three-mode model comprising the pump, signal, and idler modes provides an accurate description of the system, the experimental results are interpreted and validated through theoretical simulations. These are based on the linearization of quantum fluctuations around the steady state, a commonly adopted approach for describing quantum properties below threshold~\cite{Che16}.
Above threshold, optical parametric oscillators are instead typically described within a semiclassical framework based on the Lugiato–Lefever equations~\cite{LLe87}.

\section{Experiment}
\subsection{Setup}

A simplified version of the experimental setup is illustrated in Fig.~\ref{fig:Sec_exp}. 
As pump laser we use a CW external-cavity diode laser centered at \qty{1543.7}{nm}, that corresponds to channel 42 of the International Telecommunication Union (ITU) grid, which is the standard of modern dense wavelength-division-multiplexing (DWDM) and is tunable on about \qty{0.4}{nm} bandwidth (\qty{50}{GHz}). The measured linewidth is $\gamma_P =2\pi\times\qty{43.2(6)}{kHz}$.
The pump beam is first split into two paths. Part of the power ($95\%$) is used to generate two weak tunable probes addressing the cavity modes adjacent to the pump resonance, while the remainder ($5\%$) drives the photon-pair generation process. The latter path is amplified by a continuous-wave erbium-doped fiber amplifier (C-EDFA) to an output power of approximately \qty{200}{mW}. It then passes through an optical band pass filter (OBPF), centered at the pump wavelength to suppress amplified spontaneous emission (ASE) noise, a polarization controller and a circulator C1. It is eventually coupled into a silicon nitride resonator via a lensed fiber and an on-chip bus waveguide. On the photonic chip, the light propagating through the bus waveguide is evanescently coupled to the ring resonator mode at $\omega_0$. The resonator has a free spectral range (FSR) of around $\qty{1.1}{nm}$ ($2\pi\times\qty{142.6}{GHz}$) and an intrinsic quality factor of $\sim5\times10^{6}$. The coupling conditions are determined by the gap between the bus and ring waveguide. With a gap of \qty{500}{nm}, our system operates close to critical coupling~\cite{YAR02}, with a loaded quality factor of $\sim2\times10^{6}$, and a resulting effective cavity linewidth $\gamma$ of $\sim 2\pi \times \qty{110}{MHz}$. 
As the pump radiation is tuned closer to the ring resonance at $\omega_0$ two simultaneous effects occur in the microring. First, the intracavity power increases, enhancing the nonlinear processes. Second, thermo-optic and thermal expansion effects are induced due to heating of the microresonator~\cite{Carmon04}.

The relevant nonlinear processes include FWM, that generates signal–idler photon pairs~\cite{Bravo07}, and self- and cross-phase modulation that shifts $\omega_0$ to a new position $\omega_0^\text{NL}$. Second, the cavity resonance is further shifted mainly due to the thermo-optic effect~\cite{Arb:13, Tien:12}: heating of the ring modifies the effective refractive index, displacing the resonance frequency by an amount $\Delta\omega_0^\text{th}$. 
The instantaneous cavity resonance is then given by $\omega_c \sim \omega_0^\text{NL} + \Delta\omega_0^\text{th}$.

To account for this drift, we define an effective detuning $\delta_{\text{eff}}=\omega_p-\omega_c$ as the instantaneous frequency offset between the pump laser $\omega_p$ and the non-linearly and thermally shifted cavity resonance $\omega_c$ (see Suppl. Mat. Fig.~S1(a)). 
We continuously monitor $\delta_{\text{eff}}$ using the small fraction of the pump routed through an electro-optic phase modulator for high-order sideband generation, see Fig.~\ref{fig:Sec_exp}. The phase modulation (PM) frequency is set to \qty{35.5}{GHz}, placing the fourth-order sidebands on the adjacent cavity resonances ($\omega_0 \pm \text{FSR}$, \qty{1544.7}{nm} and \qty{1542.5}{nm}). These sideband probes are injected into the ring in the backward direction via circulator C2. By scanning the PM frequency we can determine the instantaneous FSR (also affected by temperature and non-linear effects) and $\delta_{\text{eff}}$, see Suppl. Mat. Sec.~\ref{SMSec:eff_det}.

Meanwhile, the forward-propagating output, containing both the residual pump and the FWM-generated photons, is collected from the bus waveguide via another lensed fiber. The unconverted pump component is filtered out (\qty{80}{dB} suppression) by cascading three OBPFs. Part of the light is tapped and directed to an optical spectrum analyzer (OSA) to monitor the spectral response, while the remainder is sent to a Variable Optical Attenuator (VOA) and used for the different measurements.

\begin{figure*}[t]
 \centering
 \includegraphics[width=1.\linewidth]{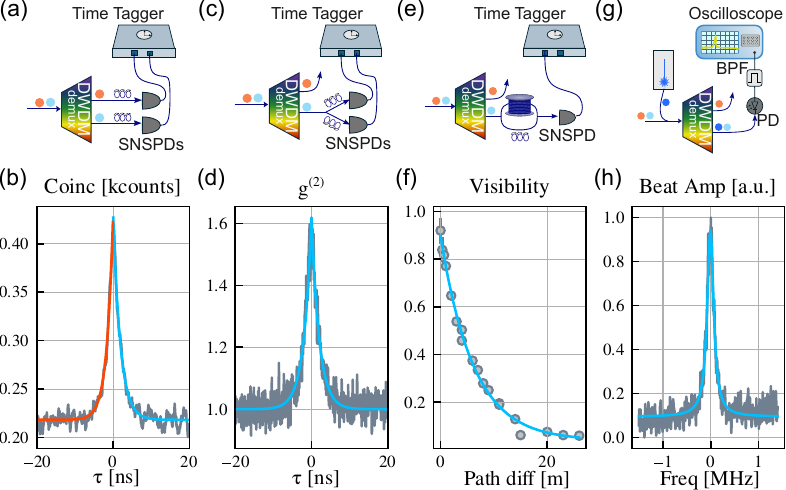}
 \caption{Simplified schematics of the measurements and representative examples. The light-blue (orange) color is used to indicate the idler (signal) mode at \qty{1530.1}{nm} (\qty{1557.3}{nm}) (a) Coincidence measurement scheme. (b) Example of coincidence measurement. Here the signal is used as start and the idler as stop on the time tagger, so that $t_\text{stop} = t_\text{start} + \tau$. The measurement shown is taken for $\delta_\mathrm{eff} = 2\pi\times\qty{154.28(3)}{MHz}$ and using the fitting function Eq.~\eqref{Eq:coinc_fit_func} we get $\tau_{-} =\qty{3.3(2)}{ns}$ and $\tau_{+} =\qty{3.7(2)}{ns}$. (c) Second-order correlation function $g^{(2)}$ scheme for the idler mode. (d) $g^{(2)}$ example data for $\delta_\mathrm{eff} = 2\pi\times\qty{151.06(6)}{MHz}$, fitted with the function Eq.~\eqref{Eq:g2_function} to obtain $\tau_\mathrm{i} = \qty{4.05(8)}{ns}$. (e) Interferometer scheme for the idler mode. (f) Interferometer measurement example. Here the path difference is set by the length of the extra fiber in one of the interferometer arms. The example data is taken for $\delta_\mathrm{eff} =2\pi\times\qty{139(1)}{MHz}$ and using the fitting function in Eq.~\eqref{Eq:visibility} we obtain $\Delta L = \qty{6.2(2)}{m}$, corresponding to $\tau_i = \qty{31(1)}{ns}$. Each data point in the figure corresponds to a single measurement, see Suppl. Mat. Sec.~\ref{SMSec:int_meas} for details. (g) Heterodyne beat scheme between the idler mode in an OPO regime~\cite{Perez23} and an ultranarrow local oscillator (LO). The LO is a tooth of an optical comb, locked to an ultrastable cavity (linewidth $<$ \qty{10}{Hz}). (h) Heterodyne data example taken for $\delta_\mathrm{eff} =2\pi\times\qty{139.3(4)}{MHz}$ and using a Lorentzian fit we obtain $\tau_i =  \qty{1.6(1)}{\mu s}$. Zero frequency corresponds to a real beat frequency of $\sim$\qty{10}{MHz}.}
 \label{fig:Sec_exp2}
\end{figure*}

Our study focuses on the photon pairs generated at the maximally phase-matched cavity modes~\cite{Zhang22}, located 12 FSRs away from the pump resonance, corresponding to wavelengths of \qty{1530.1}{nm} (ITU CH59) and \qty{1557.3}{nm} (ITU CH25) for the idler and signal modes, respectively. These are spectrally separated using a commercial DWDM filter and routed to the various measurement setups summarized in Fig.~\ref{fig:Sec_exp2}, to measure: (a) photon cross-correlation, (b) the second-order correlation function of a single optical mode, (c) interference in a variable-delay Mach–Zehnder interferometer, and (d) heterodyne beats against an ultranarrow-linewidth reference laser. 
Each of these techniques serves to characterize the first-order temporal coherence function, $g^{(1)}(\tau)$, which provides direct access to the spectral characteristics of the individual modes~\cite{Wie1930, Khi1934}.

To analyze the different measurements, we model the temporal decay of $g^{(1)}(\tau)\sim \exp{(-\tau/\tau_{s(i)})}$ as an exponential decay characterized by a coherence time, $\tau_{s(i)}$, unique to each mode. This choice is justified by the fact that the dominant contributions to decoherence arise from the pump laser phase noise and waveguide-cavity losses, which produce an approximately exponential decay of the field coherence. In the frequency domain, this is equivalent to a Lorentzian lineshape, a standard assumption in the literature~\cite{Sam19, Li25}. We revisit this assumption in Sec.~\ref{Sec:Theory}, where we derive an exact analytical solution under a mean-field approximation for the pumped mode and by neglecting sidebands' self-interactions. We demonstrate that while the Lorentzian profile remains a good approximation across our main experimental regimes, the analytical solution reveals structural deviations due to nonlinear-effect induced coupling between the modes.

Under the assumption of a purely Lorentzian line profile, the coherence time is inversely proportional to the angular-frequency FWHM (Full Width at Half Maximum) $\gamma_{s(i)}$ via:
\begin{equation}
\tau_{s(i)} = \frac{2}{\gamma_{s(i)}}.
\label{Eq:coher_times}
\end{equation}

We verify this relation across the different operating regimes.

\subsection{Methods \label{subsec:method}}
In the following we present the setup employed for each measurement. The effective detuning $\delta_{\text{eff}}$ is recorded in real time and simultaneously with each measurement, providing a consistent reference for all operating regimes.\\
The on-chip input power for all the measurements is \qty{24(2)}{mW}.

\subsubsection{Time-correlation measurements} 
 We first perform a coincidence photon-counting measurement to study quantum cross-correlations~\cite{Glauber63} in a spontaneous FWM condition.

After being separated with a DWDM demultiplexer, the idler and signal are sent to two superconducting nanowire single-photon detectors (SNSPDs), (Fig.~\ref{fig:Sec_exp2}(a)). The detection efficiency is maximized by adjusting the polarization of the incident photons. 
Counts are then recorded and precisely timestamped with a time tagger, see Suppl. Mat. Sec.~\ref{SMSec:g2}. 
For $\delta_\text{eff} = 2\pi\times\qty{154.28(3)}{MHz}$ $\sim 1.45\gamma$ the resulting coincidences histogram as a function of relative delay is shown in Fig.~\ref{fig:Sec_exp2}(b). 

\noindent Consistently with the assumed exponential model, the data are fitted by the expression:
\begin{equation}
A_+(\alpha_+ e^{-2\tau/\tau_{+}} + 1)\Theta(\tau) + A_-(\alpha_- e^{+2\tau/\tau_{-}} + 1)\Theta(-\tau),\label{Eq:coinc_fit_func}
\end{equation}
where $\Theta(\tau)$ is the Heaviside step function, $A_{+(-)}$ represents the background level from accidental coincidences, $\alpha_{+(-)}$ denotes the signal-to-noise ratio, and $\tau_{+(-)}$ is the characteristic decay time of the coincidences for positive (negative) delays.
From the fit, we find $\tau_{+} = \qty{3.7(2)}{ns}$ and $\tau_{-} = \qty{3.3(2)}{ns}$. These values are directly comparable to the cavity field lifetime $2/\gamma \sim \qty{2.8}{ns}$. Because the photon pairs are generated via spontaneous FWM within the resonator, the cavity modes act as the primary spectral filters, meaning the photon lifetime in the system is fundamentally limited by the effective cavity loss rate $\gamma$~\cite{Garay13}. Under these cavity-dominated dynamics, where the temporal envelope of both single-photon emission and pair correlation is dictated by the same cavity escape rate, the measured coincidence decay times $\tau_{+(-)}$ are equivalent to the single-mode coherence times $\tau_{s(i)}$ introduced for $g^{(1)}(\tau)$. Consequently, the inverse relationship with the linewidth holds, yielding spectral bandwidths of $\gamma_{i} = 2\pi \times \qty{85(4)}{MHz}$ and $\gamma_{s} = 2\pi \times \qty{97(5)}{MHz}$ via Eq.~\eqref{Eq:coher_times}.

The observed asymmetry in the coincidences decay time ($\tau_+ \neq \tau_-$) arises because the evanescent coupling between the microresonator and the waveguide is wavelength-dependent (dispersive)~\cite{Yariv00}, resulting in distinct effective quality factors for each mode. We confirmed this by performing weak-probe spectroscopy, which revealed a narrower linewidth of $2\pi\times\qty{112.6(9)}{MHz}$ for the cavity mode corresponding to the idler photons compared to $2\pi\times\qty{125.2(3)}{MHz}$ for the mode corresponding to signal photons.

A characteristic of photon pairs generated via spontaneous FWM is that each individual mode exhibits thermal photon-number statistics when its correlated counterpart is traced out~\cite{ALM12}. This corresponds to photon bunching, and is benchmarked by a second-order correlation function $g^{(2)}(0)$ = 2. Experimentally, $g^{(2)}(\tau)$ is estimated through an autocorrelation measurement. The accuracy of this approximation depends on the relationship between the decay time of the correlation function and the detector timing resolution, which is primarily limited by the SNSPD electronics timing jitter (typically on the order of \qty{100}{ps}). In our case, the correlation decay time is expected to be more than an order of magnitude longer than the detectors jitter. Consequently, the measured autocorrelation peak at zero delay, normalized to the uncorrelated baseline, provides a direct estimate of $g^{(2)}(0)$~\cite{Chri:2011}.

Under thermal statistics, the Siegert relation $g^{(2)}(\tau) = 1 + |g^{(1)}(\tau)|^2$ holds~\cite{Kampen81, sieg43}. 
Given the exponential decay model established for $g^{(1)}(\tau)$, the autocorrelation function for each mode takes the form:
\begin{equation}
g^{(2)}(\tau) = 1 + \beta_{s(i)} e^{-2|\tau|/\tau_{s(i)}}
\label{Eq:g2_function}
\end{equation}
where $\tau_{s(i)}$ is the single-mode coherence time defined in Eq.~\eqref{Eq:coher_times}, and $\beta_{s(i)} \le 1$ is an experimental contrast factor accounting for finite background noise or incomplete spatial filtering. For a purely spontaneous FWM process, this decay is set by the effective cavity photon lifetime ($\tau_{s(i)} \approx 2/\gamma$). 
Near the OPO threshold, however, stimulated emission begins to dominate, leading to a spectral narrowing that manifests as an increased coherence time.

To measure the $g^{(2)}(\tau)$, a single mode is isolated using the DWDM demultiplexer and analyzed with a standard Hanbury Brown–Twiss interferometer~\cite{BROWN56}, where a 50/50 fiber beam splitter directs the two outputs to separate SNSPDs, Fig.~\ref{fig:Sec_exp2}(c). The resulting $g^{(2)}(\tau)$ histogram for the idler mode, recorded at an effective detuning of $\delta_\text{eff} = 2\pi \times \qty{151.06(6)}{MHz} \sim 1.38\gamma$, is shown in Fig.~\ref{fig:Sec_exp2}(d). Fitting the data with Eq.~\eqref{Eq:g2_function} yields $\tau_{i} = \qty{4.05(8)}{ns}$, corresponding to $\gamma_{i} = 2\pi \times \qty{78(2)}{MHz}$. This value is comparable both with the effective cavity linewidth $\gamma$ and with the results obtained using the coincidences method. The shown far-detuned point has $g^{(2)}(0) = 1.63(2)$, whereas, approaching resonance, we measure $g^{(2)}(0)$ as high as $\sim 1.94$, see Suppl. Mat. Sec.~\ref{SMSec:g2}.

Notably, both correlation measurements presented here are no longer valid to evaluate the spectral linewidth as the system approaches the OPO threshold. On one hand, extracting the linewidth from the second-order autocorrelation $g^{(2)}(\tau)$ relies strictly on the validity of the Siegert relation, which breaks down the moment the emission statistic begins to deviate from a chaotic thermal state. On the other hand, while the coincidence method successfully tracks individual photon cavity lifetimes below threshold, this physical interpretation collapses above threshold. 
Here, the emission evolves from distinct spontaneously generated photon pairs into a macroscopically populated coherent field. As a result, the signal–idler coincidence histogram flattens, with ($g^{(2)}_{s,i}(\tau) \rightarrow 1$), and can no longer be used to extract the linewidth. 

\subsubsection{Phase-sensitive measurements} 
In this section we present two methods that directly probe $g^{(1)}(\tau)$ or the optical spectrum itself. Specifically, an interferometric measurement that determines $\tau_{s(i)}$, and heterodyne detection providing a direct measurement of $\gamma_{s(i)}$. These remain valid measurements across all operating regimes, as they are independent of the underlying emission statistics.

We start by introducing an interferometric measurement using a variable-delay Mach–Zehnder interferometer to characterize the first-order temporal coherence. This approach is conceptually similar to the well-known self-homodyne method used for laser linewidth characterization~\cite{chen24lin}.\\
For this measurement, we isolate one mode using the DWDM and split it on a 50/50 beam splitter. One arm is delayed by a variable fiber delay line of length $L$, while the other arm propagates without delay. The two paths are then recombined on a second 50/50 beam splitter and sent to a SNSPD followed by time-tagger electronics. The experimental setup is sketched in Fig.~\ref{fig:Sec_exp2}(e).
For a fixed difference $L$ in the interferometer's arms, the recorded counts vary over time based on the changing interference conditions induced by fluctuations in the fiber delay line, allowing us to measure the visibility of the interference as $V=(C_\text{M} -C_\text{m})/(C_\text{M}+C_\text{m})$, where $C_\text{M}$ ($C_\text{m}$) are the maximum (minimum) counts recorded on the SNSPD, see Suppl. Mat. Sec.~\ref{SMSec:int_meas}.

$V$ is expected to decrease as $L$ increases, and is related to the first-order coherence function~\cite{fox06} by
\begin{align}
 V(\tau) &= |g^{(1)}(\tau)| = V_0e^{-\tau/\tau_{s(i)}} = V_0e^{-L/\Delta L} \label{Eq:visibility} \\
 \tau_\text{s(i)} &= \Delta L n_\text{f}/c = 2/\gamma_\text{s(i)} \label{Eq:cohe_T_and_L}
\end{align}
where $\Delta L$ is the coherence length, $n_\text{f} \approx 1.5$ the refractive index in standard fibers and $V_0$ the initial visibility (in the ideal case $V_0 = V(0) = 1$).
Consistently with our assumption of a homogeneously broadened Lorentzian lineshape, the visibility envelope exhibits a purely exponential decay, directly linking the characteristic coherence length to the mode bandwidth $\gamma_{s(i)}$. Fig.~\ref{fig:Sec_exp2}(f) shows the decay of $V$ for the idler mode for $\delta_\text{eff} = 2\pi\times\qty{139(1)}{MHz}\sim 1.27\gamma$. 
For the dataset in figure we find $\gamma_i = 2/\tau_i = 2\pi\times\qty{15(1)}{MHz}$.

For sufficiently high output power levels of signal and idler modes, their bandwidths can be directly measured via heterodyne beatnote against a local oscillator (LO) field at slightly offset frequency $\omega_\text{LO}$. The beat notes are down-converted and amplified in the radio frequency (RF) domain and analyzed on an RF spectrum analyzer, see Suppl. Mat. Section~\ref{SMSec:Het}. The beat power in the RF domain is proportional to the product of the optical power of the individual beams, so employing a sufficiently strong LO enhances the detectability of a weak signal field.

As the beat linewidth corresponds to the convolution of the individual spectra of the signal (idler) and local oscillator, it is important to ensure that the spectrum of the LO field is significantly narrower than that of the mode under test. For this reason, we employ as LO a tooth of an ultrastable frequency comb that is phase-locked to a high-finesse cavity having linewidth $<$\qty{10}{Hz}. 
A schematic of the experimental apparatus is shown in Fig.~\ref{fig:Sec_exp2}(g) along with example data for $\delta_\text{eff} = 2\pi\times\qty{139.2(3)}{MHz}\sim 1.27\gamma$ in Fig.~\ref{fig:Sec_exp2}(h). For this sample we find $\gamma_i =2\pi\times\qty{193(5)}{kHz}$. We note that this bandwidth is three orders of magnitude narrower than that of the cavity and begins to approach the scale of $\gamma_p$.

\section{Theory \label{Sec:Theory}}
In the following, we adapt the open quantum system formalism to our system.
We describe the system Hamiltonian in a suitable frame, and the relevant dissipation channels. We theoretically study the stability and evolution of the photonic modes in the microring resonator by solving the Lindblad master equation for the first-order correlation function of the generated modes (signal and idler) both analytically within a linearized formalism and numerically within a cumulant expansion. 

We consider the microring resonator in the high-finesse limit where it accommodates a family of (torus-like) modes parameterized by a single integer wave number $\ell$ associated with the respective angular momentum. In particular we fix $\ell=0$ for the mode $\hat{a}_0$ of frequency $\omega_0$ that is pumped by the external pump laser (of frequency $\omega_p$) with pump detuning $\delta_0=\omega_p-\omega_0$. 
The resonant spectral frequencies $\omega_\ell = \omega_0 + \text{FSR} \,\ell + \zeta_2\,\ell^2 /2$ are written taking into account the second-order correction due to chromatic dispersion characterized by the parameter $\zeta_2>0$ (anomalous group-velocity dispersion). Higher-order contributions are neglected. Each mode, described by the bosonic operator $\hat{a}_\ell$, is assigned a uniform intrinsic loss $\gamma_{\text{in}}$ and coupling rate $\gamma_{\text{ext}}$ to an external waveguide, thereby neglecting mode asymmetry in the losses. Operating at critical coupling ($\gamma_{\text{in}} = \gamma_{\text{ext}}$), both channels contribute equally to the total bare linewidth, defined as $\gamma = 2\gamma_{\text{in}}$. In our experimental setup, we measured $\gamma =2\pi\times109.8(6)\,\text{MHz}$ and we have estimated $\zeta_2 / \gamma = 20(2) \times10^{-4}$ using finite-element simulations performed with COMSOL Multiphysics~\cite{COMSOL64}.

It is convenient to move to the rotating frame of the pump laser and the linear dispersion contributions, introducing the mode detunings $\delta_\ell =\omega_p +\text{FSR}\, \ell - \omega_\ell = \delta_0-\zeta_2 \ell^2 /2$.
Here, the parameter $\delta_0$ represents the linear ("bare") detuning of the cold cavity. We emphasize that this differs from the experimentally measured effective detuning $\delta_{\text{eff}}$, which dynamically incorporates both nonlinear and thermal resonance shifts experienced by the system under active pumping conditions.

\noindent In the rotating frame the Hamiltonian (scaled by $\hbar$) is $\hat{H}=\hat{H}_{\text{ring}}+\hat{H}_{P}+\hat{H}_{\text{int}}$~\cite{Vernon_Stronglydrivennonlinear_2015,Vernon_Quantumfrequencyconversion_2016,Pasq18,Che16}, where $\hat{H}_{\text{ring}}=-\sum_\ell \delta_\ell \hat{a}_\ell^\dagger\hat{a}_\ell$, and $\hat{H}_{P} = i \beta (\hat{a}_0^\dagger - \hat{a}_0)$, with the driving amplitude $\beta>0$ related to the input pump power $P_\text{in}$ via $\beta = \sqrt{\gamma_{\text{ext}} P_\text{in}/\hbar\omega_p}=\sqrt{\gamma P_\text{in}/2\hbar\omega_p}$.

Based on the experimentally explored regime, we only consider the efficient generation of photons at the modes $\ell=\pm\ell_{\text{exp}}$ with $\ell_{\text{exp}}=12$, denoted by idler mode $\hat{a}_i$ (with frequency $\omega_i$ and detuning $\delta_i=\delta_s$), and signal mode $\hat{a}_s$ (with frequency $\omega_s$ and detuning $\delta_s$), respectively.
These three modes (signal, the central pumped one and idler) are coupled due to Kerr interactions, $\hat{H}_{\text{int}}= \hat{H}_{\text{SPM}}+\hat{H}_{\text{XPM}}+\hat{H}_{\text{SFWM}}$, where the self-phase modulation (SPM), cross-phase modulation (XPM), and spontaneous four-wave mixing (SFWM) terms are accounted for, respectively, by
\begin{align}
 \hat{H}_{\text{SPM}} &= -\frac{g}{2} (\hat{a}_0^\dagger)^2 \hat{a}_0^2 -\frac{g}{2}\left((\hat{a}_i^\dagger)^2 \hat{a}_i^2 + (\hat{a}_s^\dagger)^2 \hat{a}_s^2 \right)\nonumber\\
 \hat{H}_{\text{XPM}} &= -2g \hat{a}_0^\dagger\hat{a}_0\left(\hat{a}_i^\dagger\hat{a}_i +\hat{a}_s^\dagger\hat{a}_s\right) -2g \hat{a}_i^\dagger\hat{a}_s^\dagger\hat{a}_i\hat{a}_s\nonumber\\
 \hat{H}_{\text{SFWM}} &= -g \hat{a}_0^2\hat{a}_i^\dagger\hat{a}_s^\dagger -g(\hat{a}_0^\dagger)^2 \hat{a}_i\hat{a}_s\,. \label{eq:inthamiltonian}
\end{align}
The latter models a process where two pump photons are converted to two photons of modes $-\ell_\text{exp}$ and $\ell_\text{exp}$. The characteristic coupling strength is $g = \hbar\omega_0^2 c n_2/(n^2V_\text{eff})$, where $n_2$ is the Kerr coefficient of the material, $n$ is the linear refractive index and $V_\text{eff}$ the effective volume of the mode. 
In our experiment we estimated $g/\gamma=2.8(3)\times10^{-9}$.

We describe the dynamics of the ring modes by a density matrix $\varrho$ obeying the Lindblad master equation 
\begin{align}
\label{eq:Lindblad}
\frac{d}{dt}\varrho 
= 
-i [\hat{H},\varrho] + \gamma\,\sum_{\ell}\mathcal{D}[\hat{a}_\ell](\varrho),
\end{align}
where the dissipator is defined by $\mathcal{D}[\hat{a}_\ell](\varrho) = \hat{a}_\ell \varrho\, \hat{a}_\ell^\dagger - \frac{1}{2} (\hat{a}_\ell^\dagger \hat{a}_\ell \varrho + \varrho\, \hat{a}_\ell^\dagger \hat{a}_\ell)$.
We aim to model signal-idler pairs generation due to SFWM before the onset of the OPO. In this regime, we first consider a mean-field approximation to exploit the fact that the occupation of the sidebands $n_{s/i}$ is negligible with respect to the occupation of the pumped mode, extending the analysis of Ref.~\cite{Vernon_Stronglydrivennonlinear_2015}. 
We also choose the initial state $|\alpha_0(0)\rangle$ for the pumped mode as a coherent state for which $\hat{a}_0|\alpha_0(0)\rangle=\alpha_0(0)|\alpha_0(0)\rangle$. 
 We can then take the semiclassical approximation in which $\hat{a}_0\simeq \alpha_0 (t)\equiv \langle \hat{a}_0(t)\rangle = \text{Tr}(\hat{a}_0 \varrho(t))$ and consider the undepleted pump limit where we assume the pumped mode is unaffected by the XPM and SFWM interaction terms involving $s$ and $i$. 

\begin{figure}
 \centering
 \includegraphics[width=1\linewidth]{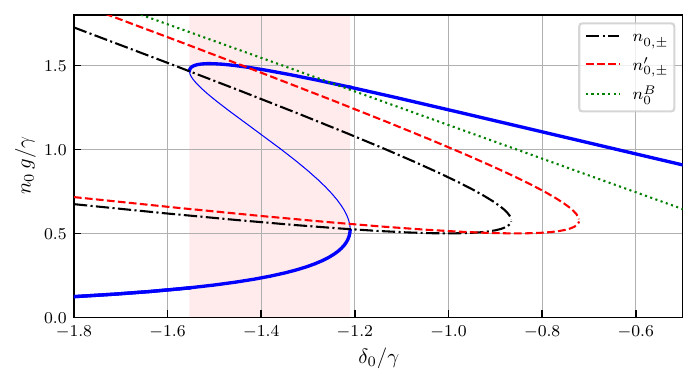}
 \caption{Scaled pumped-mode occupation $gn_0/\gamma$ (solid thick curves: stable; thin curve: unstable) for $P_\text{in}=\qty{24}{mW}$, and stability of the pump and sideband modes, as a function of the cold-cavity detuning $\delta_0/\gamma$. We show the threshold occupancy of mode 0, associated with the onset of instability for $\delta\hat a_0$ ($n_{0,\pm}$, black dot-dashed curve), and for $\hat a_{s/i}$ ($n_{0,\pm}^\prime$, red-dashed curve). We operate in the bistability regime (colored band) at the high-occupation stable solution. $n^B_0$ (dotted curve) is the line for the bifurcation of the signal-idler mode ($|l| = 12$). Where this boundary crosses the power-dependent stability curve, it separates the split-spectrum and gain-narrowing regimes.}
 \label{fig:stability}
\end{figure}

The Lindblad master equation for the pumped mode then reduces to an equation of motion for the coherent field amplitude $\alpha_0(t)$
\begin{equation}
\label{eq:mode0EoM}
 \left[\frac{d}{dt} +\frac{\gamma}{2}-i(\delta_0+{g} |\alpha_0(t)|^2)\right]\alpha_0(t) = \beta\,.
\end{equation}
The average occupation ${n_0}=|\alpha_0|^2$ in the stationary state is a solution to the following cubic equation 
\begin{align}\label{eq:bistabilityn0}
 C({n_0}) \equiv \left[(\delta_0+{g} {n_0})^2+ \left(\frac{\gamma}{2}\right)^2\right]{n_0} -\beta^2=0\,,
\end{align}
with the amplitude being $\alpha_0 = \beta /(\gamma/2 -i(\delta_0+{g} {n_0}) )$.
Solutions to Eq.~\eqref{eq:bistabilityn0} (see Fig.~\ref{fig:stability}) define a classical background for the dynamics of the sidebands. 
The stability of the fluctuations $\delta\hat a_0$ is determined by the condition $\partial C/\partial {n_0} > 0$~\cite{Vernon_Stronglydrivennonlinear_2015}, which is satisfied for $n_0<n_{0,-}$ or $n_0>n_{0,+}$, with $n_{0,\pm}=[-2\delta_0\pm(\delta_0^2-\delta_c^2)^{1/2}]/3g$ and $\delta_c = -\sqrt{3}\gamma/2$ (dot-dashed line). 
 We operate above the pump-bistability critical drive $\beta> \gamma^{3/2}/(3^{3/2}g)^{1/2}$, which in our experiment is around $\qty{12}{mW}$ (see Suppl. Mat.~\cite{OurSM}). This means that, for sufficiently low pump laser frequency, we reach a regime of bistability (vertical band) with two stable real solutions and one real unstable solution for $n_0$ (solid thick and thin lines, respectively). In Fig.~\ref{fig:stability}, we consider the experimental value $P_\text{in}=\qty{24}{mW}$. 

The limit $n_0\gg n_{s/i}$ allows for the linearization of the dynamics of the sidebands in Eq.~\eqref{eq:Lindblad} by keeping terms only up to first order in the quantum operators $\hat a_{s/i}$, so that the equation of motion for $\langle\hat a_{s/i}\rangle$ can be written as
\begin{equation}
	\label{eq:mat_stability_sidebands}
		\frac{d}{dt} \begin{pmatrix}
				 \langle \hat{a}_s(t)\rangle\\
				 \langle \hat{a}_i^\dagger(t)\rangle
				 \end{pmatrix} = \begin{pmatrix}
				 -\frac{\gamma}{2}+i(\delta_s+2{g} {n_0}) & i {g} \alpha_0^2\\
				 -i {g} (\alpha_0^\ast)^2 & -\frac{\gamma}{2}-i(\delta_s+2{g} {n_0})
				 \end{pmatrix} \begin{pmatrix}
				 \langle \hat{a}_s(t)\rangle\\
				 \langle \hat{a}_i^\dagger(t)\rangle
				 \end{pmatrix}\,,
\end{equation}
analogous to the one derived in Ref.~\cite{Vernon_Stronglydrivennonlinear_2015}, with the notable difference that dispersion makes the sideband detuning differ from the pumped-mode detuning, so that the signal-idler detuning $\delta_s$ replaces $\delta_0$. 
For the system to remain in the sub-OPO regime, the solutions to Eq.~\eqref{eq:mat_stability_sidebands} must decay over time. This requires the real part of both eigenvalues of the above matrix $-\gamma/2 \pm \bar{\rho}$, with $\bar{\rho} = \sqrt{{g}^2{n_0}^2-(\delta_s+2{g} {n_0})^2}$, to be negative. This yields the sidebands' stability condition
\begin{equation}
\label{eq:stability_sidebands}
 \Delta\equiv\left(\frac{\gamma}{2}\right)^2-\bar{\rho}^2 >0\;,
\end{equation}
which is satisfied for $n_0<n_{0,-}^\prime$ or $n_0>n_{0,+}^\prime$ with $n_{0,\pm}^\prime=[-2\delta_s\pm(\delta_s^2-\delta_c^2)^{1/2}]/3g$. The intersection between these stability bounds and the stable $n_0$ occupancies (crossing of dashed and solid lines in Fig.~\ref{fig:stability}) occurs at detunings depending on the combination $\beta^2g/\gamma^3$~\cite{Vernon_Stronglydrivennonlinear_2015,Chembo_Modalexpansionapproach_2010}.
This threshold corresponds to the onset of instability for the sidebands and the system entering the OPO region.
In the experimental procedure, we approach this bistability regime by scanning from higher to lower detunings, resulting in a continuous increase of the pumped mode occupancy $n_0$.

The sideband dynamics is conveniently described by the time evolution of the mean values of the total sideband occupation $\hat{K} =\frac12 (\hat{a}_s^\dagger\hat{a}_s +\hat{a}_i^\dagger\hat{a}_i +1)$, and the pair-creation operator $\hat{K}_+ = \hat{a}_s^\dagger\hat{a}_i^\dagger$, with its Hermitian conjugate $\hat{K}_- = \hat{a}_s\hat{a}_i$ (see Suppl. Mat.~\cite{OurSM} for the rationale underlying this choice and the corresponding equations of motion). For the stationary state: 
$ \langle \hat{K}\rangle = ((\gamma/2)^2 + (\delta_s+2g n_0)^2)/(2\Delta)$ and $\langle \hat{K}_+\rangle =\langle \hat{K}_-\rangle^* = -i g (\alpha_0^*)^2 (\gamma/2 -i(\delta_s+2g n_0)) / (2\Delta)$.
$\Delta$ is therefore the quantity characterizing the behavior of the sidebands' occupation with varying detuning and in particular it gives an estimation of the relative strength of XPM and SFWM contributions. 
The sideband steady state depends on the pumped mode occupation $n_0$. 
In particular, for large detunings, $\bar{\rho}$ is imaginary due to XPM induced shift of spectral lines, and the growth of $\langle \hat{K}\rangle$ is mainly driven by XPM between the pumped mode and the sidebands.
Here the factor $1/\Delta \sim 1/((\gamma/2)^2+\delta_s^2)$, which makes the variation of $\langle \hat{K}\rangle$ with detuning small.
As we enter the pumped mode's bistability regime, $\langle \hat{K}\rangle$ has more than one solution. 
Within this regime, $\bar{\rho}$ eventually becomes real when SFWM-induced pair generation becomes dominant, and the growth of $\langle \hat{K}\rangle$ is enhanced by the increasingly small quantity $\Delta$. 
In particular, as $\bar{\rho}$ approaches $\gamma/2$, $\langle \hat{K}\rangle$ formally diverges, corresponding to the onset of OPO. This unphysical prediction occurs because, near the threshold detuning, signal-idler pairs are generated faster than they are removed from the ring. As a result, the system cannot reach a steady state within our model. In particular, pump depletion and pure sideband self- and cross-phase modulation can no longer be neglected; additional resonator modes may also become relevant. This means that we are outside the range of validity for the considered linearized model. 

The emission spectrum $\nu_s$ in the stationary state for the sidebands is defined in terms of the unnormalized first-order correlation function $G^{(1)}$ as $\nu_s(\omega) = \int d\tau G^{(1)}(\tau)e^{i\omega\tau}$~\cite{Khi1934, Wie1930} (see Suppl. Mat.~\cite{OurSM} for derivation of $G^{(1)}$).
We find the spectrum
\begin{align}
 \nu_s(\omega) =&\frac{g^{2} n_0^2}{\Delta}\Re \left(\frac{{\gamma}-i \omega}{(\frac{\gamma}{2}-i \omega)^2-\bar{\rho}^2}\right)
 = \frac{g^{2} \gamma n_0^2}{|\frac{\gamma}{2}-\bar{\rho}+i \omega|^2|\frac{\gamma}{2}+\bar{\rho}+i \omega|^2}. \label{eq:spectrum_th} 
\end{align}
This expression takes the form of a product of two Lorentzians. 
When $\bar{\rho}$ is imaginary, $\bar{\rho}=i\Omega$, $\Omega\in \mathbb{R}$, these Lorentzians have identical full widths $\gamma_s = \gamma$ and are centered on $\omega=\pm \Omega$. For large detunings, $\bar{\rho} \approx i|\delta_s|$, the spectrum is peaked at $\omega = \pm|\delta_s|$. This is a consequence of the trade-off between energy conservation and enhancement of the pair generation process.
As $\delta_s$ decreases, $n_0$ increases and the extent of the splitting $2\Omega$ is reduced due to XPM effectively countering the pump detuning. This splitting eventually vanishes at the bifurcation boundary $n_0=n_0^B=- \delta_s/g$, marking the regime where SFWM-induced signal-idler pairs generation dominates. 
If the detuning is further decreased, $\bar{\rho}$ becomes real and stops contributing to spectral splitting, instead contributing to the linewidth. The two Lorentzians are then both centered on $\omega= 0$, with respective full widths $\gamma_\pm = \gamma \pm 2\bar{\rho}$. As $\bar{\rho}$ approaches $\gamma/2$ near the OPO-threshold, the smaller of these two widths becomes dominant, leading to a line shape with overall effective width $\gamma_s=\gamma-2\bar{\rho}$, which formally vanishes at the OPO threshold.

The bifurcation boundary $n_0^B$ is plotted in Fig.~\ref{fig:stability} as a dotted line. The intersection of this universal line with the power-dependent stability curve (solid blue) determines the case-specific detuning for which the spectral splitting collapses and the subsequent linewidth narrowing begins.

To conclude, we note that, just like for the sidebands, we can define a pump-mode fluctuations parameter $\bar{\rho}_0 = \sqrt{g^2 n_0^2 - (\delta_0 + 2g n_0)^2}$. Unlike the sidebands, where dispersion can drive $\bar{\rho}$ to become real, triggering parametric gain, the pump parameter $\bar{\rho}_0$ remains strictly imaginary across all stable operating regimes. The quantity $\Im(\bar{\rho}_0)$ physically represents the nonlinearly shifted detuning of the pumped mode quantum fluctuations, stemming from the SPM interaction. Apart from negligible second-order dispersion, this is equivalent to the shifted detuning of the $|l|=1$ modes. Thus, the experimental measurement of $\delta_\text{eff}$ corresponds to $\Im\bar{\rho}_0$, with the exception of slow thermal resonance shifts.

\section{Results}

We experimentally investigated the linewidth of the emitted idler photons in proximity of the transition from the SFWM to the OPO regime. This linewidth is characterized as a function of the experimental effective detuning $\delta_\text{eff}$, which inherently includes both thermal and nonlinear shifts of the pumped resonator mode. To gain insight into the underlying dynamics of the transition, we compare these measurements to a theoretical simulation. In our model, we simulate the linewidth as a function of the purely Kerr-induced shift, $\Im(\bar{\rho}_0)$. Indeed, a complete quantitative modeling of the concurrent thermal dynamics would require an exhaustive parametrization of localized thermal dissipation coefficients. However, while thermal dynamics induce a power-dependent resonance shift and a deformation of the effective detuning profile, these effects primarily reparameterize~\cite{Herr26} the detuning axis rather than alter the core physics. Thus, focusing strictly on the electronic Kerr nonlinearity captures the essential physics of the threshold behavior, albeit with a shifted or compressed detuning scale.

As expected, we show here that the SFWM-to-OPO transition exhibits a clear threshold behavior corresponding to the fact that it requires the intracavity pump power to be high enough for the intensity-dependent parametric generation rate (proportional to the parameter $\bar{\rho}$) to overcome the total loss rate $\gamma/2$. Once this condition is met, the signal–idler photon pairs generated spontaneously act as seeds for stimulated amplification, leading to coherent oscillation and enhanced output power~\cite{Pasq18, Kipp18, Kipp04}.

\begin{figure}[t]
 \centering
 \includegraphics[width=1\linewidth]{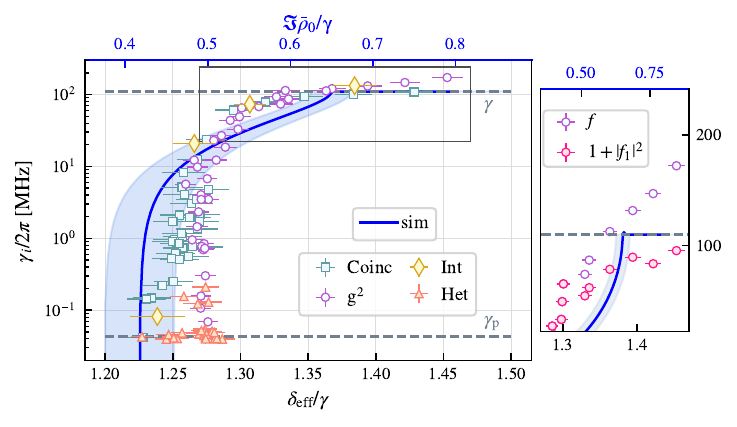}
 \caption{(Main, Bottom axis) Idler mode linewidth as a function of normalized detuning $\delta_\text{eff}/\gamma$, with $\gamma = 2\pi \times 109.8(6)\,$MHz. Experimental linewidths extracted from measurements of temporal coincidences (light-blue squares), autocorrelation (purple circles), variable visibility in a Mach-Zehnder interferometer (yellow diamonds), and heterodyne beat (orange triangles). The lower (upper) gray dotted line indicates the linewidths of the pump laser (the cavity mode at the pump wavelength). 
 \\
 (Main, Top axis) Numerical simulation of idler mode linewidth as a function of the normalized Kerr-shifted detuning $\Im{(\bar{\rho}_0)}/\gamma$. The blue solid line corresponds to an input power of \qty{24}{mW}, while the light-blue shaded area corresponds to a power uncertainty of $\pm$\qty{2}{mW}.
 \\
 (Inset) Comparison of the autocorrelation data fitted using two different models: a single Lorentzian ($g^{(2)}(\tau)$ fitted by $f$ given in Eq.~\eqref{Eq:g2_function}) and two Lorentzians of same width and separation $2\Omega$. For the latter $g^{(1)}$ is given by $f_1$ in Eq.~\eqref{Eq:fit_function_g1}, and the fitting function is obtained assuming that $g^{(2)} = 1 + |g^{(1)}|^2$ holds.}
 \label{fig:Fig4}
\end{figure}

In our experiment, the input pump power is chosen to permit scanning across this transition. Starting from large detunings where the intracavity power is low, we decrease the detuning until the circulating power reaches the threshold required to enter the OPO regime. Hence, $\delta_\text{eff}$ serves as the parameter that drives the system from a regime dominated by spontaneous emission to one characterized by stimulated emission.
Experimentally, we observe the onset of parametric oscillation at an effective detuning of $\delta_\text{eff} \sim 1.25\gamma$.

As shown in Fig.~\ref{fig:stability}, the input power used in our measurements (\qty{24}{mW}) exceeds the minimum power of \qty{12}{mW} for pumped-mode bistability. This higher experimental pump power relaxes the requirement for cavity resonance enhancement, thereby shifting the OPO threshold toward larger detuning absolute values compared to the minimum-power case.

We exploit the methods described in Sec.~\ref{subsec:method} to measure the linewidth of the generated mode while the pump laser temperature is tuned so that its frequency is decreased from a blue-detuned condition toward the thermally-shifted cavity resonance. For each regime, we employ at least two measurement methods for comparison.
The results are summarized in Fig.~\ref{fig:Fig4}. 

The light-blue squares represent the results of the coincidence measurements between the signal and idler photons. As previously discussed, the asymmetry between $\gamma_i$ and $\gamma_s$ is attributed to a wavelength-dependent $\gamma$, so here we report only the results for the idler mode. For large effective detuning ($\delta_\text{eff} \gg 1.25\gamma$), the linewidth of the emitted photons approaches the effective cavity linewidth. However, as $\delta_\text{eff} \rightarrow 1.25\gamma$, the bandwidth narrows. This narrowing is accompanied by a decrease in the overall coincidence counts; as the system approaches the OPO threshold, temporal cross-correlations diminish, signaling the transition from discrete photon pair emission to continuous coherent oscillation.

The estimate of $\gamma_i$ obtained from the second-order correlation function measurement (purple circles) gives consistent results, reproducing the same narrowing trend of the linewidth. Close to the OPO threshold, this method similarly encounters limitations; as the system transitions toward a coherent state, the diminishing non-classical correlations render the second-order correlation measurement unreliable.

Therefore, heterodyne measurements are employed to extend our characterization into the OPO regime, where single-photon–regime correlation techniques fail, see orange triangles in Fig.~\ref{fig:Fig4}. Each heterodyne trace corresponds to an average over six consecutive measurements. We concentrate the acquisition in the vicinity of the transition, where the system is most sensitive to drift and threshold fluctuations: in this region it is experimentally challenging to measure both second-order time-correlation functions and to extract a meaningful heterodyne signal. As expected at the threshold, the measurement is noisy; however, the inferred linewidths show good agreement with those obtained from the photon-correlation analysis below threshold.
Once the transition into the OPO regime has occurred, signaled experimentally by the sudden appearance of well-resolved signal and idler sidebands on the optical spectrum analyzer, together with the flattening of both the autocorrelation $g^{(2)}(\tau)$ and the signal–idler coincidence traces, the heterodyne measurement becomes the only reliable method to extract the linewidth. In this regime, the linewidth is no longer governed by the effective cavity decay rate $\gamma$, but mainly influenced by pump phase noise as well as by nonlinear Kerr-induced effects, such as self- and cross-phase modulation, which convert intensity fluctuations into phase noise~\cite{Mats:07, DelH:11}. Though active across both regimes, this intensity-to-phase noise conversion is strongly amplified in the OPO state by the significantly higher parametric generation rate.
 
As a final verification, we include four data points obtained from an interferometric measurement (yellow diamonds). Unlike the previous techniques, this method probes the stability of the system indirectly, since acquiring each point requires manually scanning the optical path difference of the Mach–Zehnder interferometer over a relatively long time. Despite this additional sensitivity to slow drifts, the interferometric points agree well with the overall trend across all operating regimes. Furthermore, even for small $\delta_\text{eff}$ we find that the interferometric measurements exhibit an exponential decay of the coherence, see Suppl. Mat.~\cite{OurSM}, 
consistent with the assumption in Eq.~\eqref{Eq:visibility} of a Lorentzian lineshape. This indicates that on a sub-hour timescale, the linewidth is not dominated by thermal fluctuations, which would give a Gaussian lineshape, but is instead primarily determined by pump-induced phase diffusion transferred to the signal and idler fields.

The variability between the different measurements is mainly associated with slow fluctuations of the EDFA gain. On a daily timescale, these fluctuations result in variations of up to \qty{2}{mW} in the on-chip pump power, likely due to slow thermal effects within the amplifier. Since these variations occur over several hours, they lead to changes in $P_\text{in}$ between successive data acquisitions, ultimately manifesting as a horizontal shift in $\delta_\text{eff}$ from one measurement to the next.

The analysis presented up to this point relies on a single exponential fit, the standard approximation in the literature, which corresponds to a simple Lorentzian spectrum. However, scanning $\delta_\text{eff}$ tunes the system parameter $\bar{\rho}$ from the imaginary to the real domain. As described by Eq.~\eqref{eq:spectrum_th}, this represents a transition from a spectrum characterized by two Lorentzians with a frequency splitting $2\Omega$ and uniform full width $\gamma$, to a regime of two Lorentzians both centered at zero but with distinct linewidths $\gamma_{\pm}$. Rather than a simple exponential decay, the correct fitting function is then
\begin{align}
f_{\text{fit}} =
\begin{cases}
f_1 \sim A_0 e^{- \tau \gamma / 2} \cos(\Omega\tau + \phi) & \text{for imaginary } \bar{\rho} \\
f_2 \sim A_0 \left( \frac{e^{-\tau \gamma_-/2}}{\gamma_-} - \frac{e^{-\tau \gamma_+/2}}{\gamma_+} \right) & \text{for real } \bar{\rho}, 
\end{cases}
\label{Eq:fit_function_g1}
\end{align}
with $A_0$ amplitude and $\phi$ initial phase (see Suppl. Mat.~\cite{OurSM} for derivation). 
 
In the first regime, a single exponential fit tends to overestimate the linewidth, as illustrated in the zoomed region of Fig.~\ref{fig:Fig4}. While the exponential fit suggests a continuous growth in linewidth, the correct fitting function reveals the expected saturation towards $\gamma$.
However, as $\delta_{\text{eff}}$ is reduced toward the OPO threshold ($\bar{\rho} \to \gamma/2$), $\gamma_{-}$ approaches zero while $\gamma_{+}$ approaches $2\gamma$. In this limit, the $f_2$ function converges to a single Lorentzian behavior where the decay is dominated by the narrowest component, $\gamma_{-} \sim \gamma_\text{s,i}$.
We highlight that within the real $\bar{\rho}$ regime, the pair-generation rate increases sharply as the system approaches the stimulated emission threshold. Consequently, we expect the $\gamma_{-}$ term to quickly become the dominant contribution.

Experimentally, however, the signal-to-noise ratio in our current acquisition does not permit a clear distinction between the two fitting regimes. We find that the linewidth $ \gamma_{-}$ extracted from the double-exponential model is compatible with the single exponential estimate $\gamma_\text{s,i}$ within the fitting uncertainties (see Suppl. Mat.~\cite{OurSM} for a detailed comparison). Consequently, we maintain the single exponential approach for the broad parameter scan, as it provides a reliable estimate of the dominant decay scale without over-parameterizing the fit.

We compare the experimental data with numerical simulations of the linewidth as a function of $\Im{\bar{\rho}_0}$. As discussed in Sec.~\ref{Sec:Theory}, the linearized theory presented fails very close to the OPO-threshold, due to the increasing population in the sideband modes. While accurately describing this regime requires lifting the assumption of undepleted pump, which is beyond the scope of this work, including the SPM and XPM terms involving only signal and idler modes improves the description of the dynamics close to threshold. We incorporate these terms in a cumulant expansion approach, which offers a systematic way of expanding averages of higher-order operator products in terms of products of lower order expectation values, up to a chosen truncation order. We employed the QuantumCumulants.jl Julia library~\cite{Plankensteiner_QuantumCumulantsjlJuliaframework_2022} to numerically solve the equations of motion derived from Eq.~\eqref{eq:Lindblad} for $n_s$ and $g^{(1)}$. We use the semiclassical approximation $\hat{a}_0\to\alpha_0$ for the pump mode, where $\alpha_0$ is the stationary high-occupation stable solution of Eq.~\eqref{eq:bistabilityn0}. 
Finally, we apply the Wiener–Khintchine theorem~\cite{Wie1930, Khi1934} to derive the emission spectrum, from which the linewidth is extracted by fitting with Eq.~\eqref{Eq:fit_function_g1} (see Suppl. Mat. Sec.~\ref{SMSec:fitg1}).

The extracted theoretical linewidth for $P_\text{in} = 24\text{ mW}$ is shown by the blue solid line in Fig.~\ref{fig:Fig4}. The accompanying light-blue shaded region represents the uncertainty bounds corresponding solely to variations in the input power ($P_\text{in} = 24 \pm 2\text{ mW}$), consistent with the experimentally observed fluctuations. The pronounced broadening of the uncertainty band in the vicinity of the OPO threshold reflects the increased sensitivity of the linewidth to small pump-power variations in this regime, whereas the dependence becomes significantly weaker at larger detunings. The resulting uncertainty closely matches the experimental scatter, indicating that the observed variability is largely accounted for by the measured pump-power fluctuations.

Apart from the precise location of the OPO threshold, both experiment and theory successfully capture the same linewidth-narrowing trend, with both curves asymptotically approaching the resonator-dominated linewidth at large detunings. In contrast to the experimental data, however, the simulation predicts an unconstrained narrowing of the linewidth toward zero at the threshold. This difference stems from the fact that our model does not incorporate technical noise on the pump laser, which establishes a physical floor in the experiment.
Furthermore, these noise-free simulations exhibit a cusp in the linewidth, where its first derivative is not continuous as $\bar{\rho}$ transitions into the real domain, consistent with the square-root dependence governing the eigenvalue splitting. This abrupt feature is smoothed out in the experimental data due to technical noise and the resolution limits of the fitting procedures detailed above. Finally, within the $\bar{\rho} \in \mathbb{R}$ regime, the simulations clearly resolve the individual decay rates $\gamma_+$ and $\gamma_-$. This clear resolution confirms that $\gamma_-$ rapidly becomes the dominant spectral contribution, fully justifying our choice to plot $\gamma_i \sim \gamma_-$ in Fig.~\ref{fig:Fig4}. 
As a consistency check, we verified that the threshold position predicted by the second- and fourth-order cumulant expansions for the emission spectrum $\nu_s(\omega)$ agrees to within a few percent, confirming the convergence of our expansion method within our experimental parameter uncertainties.

As a final remark, we note that the thermal control of our platform is implemented macroscopically via a single Peltier module situated beneath the chip-and-holder assembly. While this setup successfully mitigates slow drifts stemming from ambient room temperature variations, it cannot actively compensate for fast thermo-optic dynamics, which we monitor through $\delta_\text{eff}$. However, advanced designs utilizing integrated local microheaters for precise, single-resonator stabilization~\cite{PerLo:21, Josh16}, or, more recently, copper-free wafer architectures that are inherently less sensitive to thermo-optic effects~\cite{Ji25}, are documented in the literature. Consequently, we deliberately avoided an explicit thermal treatment in our model, as our simplified approach successfully isolates the fundamental quantum-optical dynamics without introducing platform-specific thermal engineering complexities.
This simplification is physically justified because the primary thermo-optic effect manifests as a dispersive shift and transition-shape deformation, which does not modify the parametric coupling $g$ to first order (see Suppl. Mat. Sec.~\ref{SM:gvsT}).

\section{Conclusions}
In summary, we have demonstrated a robust experimental framework to precisely characterize the spectral evolution of four-wave mixing across the transition from the spontaneous to the stimulated regime. By integrating four complementary techniques: temporal coincidences, autocorrelation, Mach–Zehnder interferometry, and heterodyne spectroscopy, we monitored a continuous linewidth narrowing of more than three orders of magnitude, providing a comprehensive mapping of emission coherence that remains inaccessible via single-modality measurements.

Coincidence measurements, in particular, reveal a distinct linewidth asymmetry between signal and idler photons, which we attribute to wavelength-dependent variations in the loaded quality factors of the microresonator.

To accurately control and track the device operating point, we measure in real-time the effective detuning between the pump laser and the thermally shifted fluctuations of the pumped mode, $\delta_{\text{eff}}$. This allowed us to map the continuous evolution from spontaneous photon-pair emission to coherent field buildup as a function of $\delta_{\text{eff}}$, providing direct insight into the threshold dynamics of the system and establishing a clear path to map these dynamics across distinct physical regimes.

For large detuning, well below the optical parametric oscillation (OPO) threshold, the emission is dominated by spontaneous processes and the linewidth is set by the hundred-MHz-level resonator bandwidth, which filters the broadband nonlinear emission. This regime is particularly attractive for applications such as quantum key distribution. As the pump approaches the parametric oscillation threshold, stimulated emission progressively enhances coherence, driving a continuous spectral narrowing toward the pump-linewidth scale. While this effect has been predicted theoretically, we refine the description by incorporating chromatic dispersion to identify the instability point, finding good agreement with our experimental observations across the transition. Numerical simulations were conducted using the QuantumCumulants.jl library, which provides a scalable framework for the systematic inclusion of higher-order correlations, an essential step for capturing the system's behavior beyond the linearized regime.

At the boundary of the model's validity, corresponding to the OPO threshold, the linearized theory predicts zero-linewidth modes. In practice, however, the system enters a phase-diffusion regime where the coherence is primarily limited by the phase noise of the pump laser, not included in the current theoretical framework. 

Finally, we evaluate the appropriate functions to fit the temporal decay of the first-order correlation function across all regimes. By comparing our results to the standard single-exponential approximation, we show that while this model overestimates the linewidth far from threshold, it becomes increasingly accurate as the system approaches OPO. In this limit, the transition toward a Lorentzian lineshape and the influence of the experimental noise floor make the single-exponential fit a reliable and practical tool for linewidth estimation. 

Our results deepen the understanding of microresonators as sources of spectrally tunable quantum light. By bridging the gap between theory and experiment, this work provides a practical tool for designing on-chip quantum sources with tailored spectral properties and a validated methodology for their precise characterization across all operational regimes.

\begin{backmatter}
\bmsection{Acknowledgment}
The authors thank Marco Liscidini (University of Pavia) for useful discussions, and Elio Bertacco (INRiM) for technical support. 
The results presented in this article had been achieved in the context of the following projects:
QUID (QUantum Italy Deployment), which is funded by the European Commission in the Digital Europe Programme under the grant agreements number 101091408 and 101091561;
QU-TEST, which had received funding from the European Union's Horizon Europe under the grant agreement number 101113901; 
DigiMiQ (INFRA+, Piemonte Region);
QCIMED project, co-funded by the European Union under the Connecting Europe Facility (CEF Digital), Grant Agreement No. 101249740;
CIPHER project, co-funded by the European Union under Grant Agreement No. 101304918;
the Project No. G6026 SPS NATO;
QUAQK, from the Italian Ministero dell’Università e della Ricerca (project PRIN-2022KH2KMT QUAQK);
WHITECH, funded by Italian Space Agency.

\bmsection{Supplemental document}
See supplemental document for supporting content.

\end{backmatter}

\bibliography{bibliography}

@PREAMBLE{
 "\providecommand{\noopsort}[1]{}" 
 # "\providecommand{\singleletter}[1]{#1}%" 
}

@unpublished{GiaccariOPO2026,
  author = {Giaccari et al., Stefano},
  note = {in preparation},
  year=2026
}

@article{Vernon_Quantumfrequencyconversion_2016,
  title = {Quantum Frequency Conversion and Strong Coupling of Photonic Modes Using Four-Wave Mixing in Integrated Microresonators},
  author = {Vernon, Z. and Liscidini, M. and Sipe, J. E.},
  year = 2016,
  month = aug,
  journal = {Phys. Rev. A},
  volume = {94},
  number = {2},
  pages = {023810},
  publisher = {American Physical Society},
  doi = {10.1103/PhysRevA.94.023810},
  note    = {\url{https://doi.org/10.1103/PhysRevA.94.023810}}
}

@article{Plankensteiner_QuantumCumulantsjlJuliaframework_2022,
  title = {{{QuantumCumulants}}.Jl: {{A Julia}} Framework for Generalized Mean-Field Equations in Open Quantum Systems},
  author = {Plankensteiner, David and Hotter, Christoph and Ritsch, Helmut},
  year = 2022,
  month = jan,
  journal = {Quantum},
  volume = {6},
  pages = {617},
  publisher = {Verein zur F\"orderung des Open Access Publizierens in den Quantenwissenschaften},
  doi = {10.22331/q-2022-01-04-617},
  note    = {\url{https://doi.org/10.22331/q-2022-01-04-617}},
}

@article{Chiribella_ApplicationsgroupSU1_2006,
  title = {Applications of the Group {{SU}}(1, 1) for Quantum Computation and Tomography},
  author = {Chiribella, G. and D'Ariano, G. M. and Perinotti, P.},
  year = 2006,
  month = nov,
  journal = {Laser Phys.},
  volume = {16},
  number = {11},
  pages = {1572--1581},
  issn = {1555-6611},
  doi = {10.1134/S1054660X06110119},
  langid = {english},
  note    = {\url{https://doi.org/10.1134/S1054660X06110119}}
}

@article{wal15,
  title={Quantum optics: Science and technology in a new light},
  author={Walmsley, IA},
  journal={Science},
  volume={348},
  number={6234},
  pages={525--530},
  year={2015},
  publisher={American Association for the Advancement of Science},
  note    = {\url{https://doi.org/10.1126/science.aab0097}},
}

@article{Vernon_Stronglydrivennonlinear_2015,
  title = {Strongly Driven Nonlinear Quantum Optics in Microring Resonators},
  author = {Vernon, Z. and Sipe, J. E.},
  year = 2015,
  month = sep,
  journal = {Phys. Rev. A},
  volume = {92},
  number = {3},
  pages = {033840},
  publisher = {American Physical Society},
  doi = {10.1103/PhysRevA.92.033840},
  note    = {\url{https://doi.org/10.1103/PhysRevA.92.033840}},
}

@article{Cas17,
  title = {Integrated sources of photon quantum states based on nonlinear optics},
  author = {Caspani, Lucia and Xiong, Chunle and Eggleton, Benjamin J. and Bajoni, Daniele and Liscidini, Marco and Galli, Matteo and Morandotti, Roberto and Moss, David J.},
  journal = {Light Sci. Appl.},
  volume = {6},
  number = {11},
  pages = {e17100},
  year = {2017},
  doi = {10.1038/lsa.2017.100},
  note = {\url{https://doi.org/10.1038/lsa.2017.100}}
}

@article{Bra21,
author = {\v{S}imon Br\"{a}uer and Petr Marek},
journal = {Opt. Express},
number = {14},
pages = {22648--22658},
publisher = {Optica Publishing Group},
title = {Generation of quantum states with nonlinear squeezing by Kerr nonlinearity},
volume = {29},
month = {Jul},
year = {2021},
note = {\url{https://opg.optica.org/oe/abstract.cfm?URI=oe-29-14-22648}},
doi = {10.1364/OE.427637},
}

@article{Fek2013,
  title = {Ultranuarrow-band photon-pair source compatible with solid state quantum memories and telecommunication networks},
  author = {Fekete, Julia and Riel{\"a}nder, Daniel and Cristiani, Matteo and de Riedmatten, Hugues},
  journal = {Phys. Rev. Lett.},
  volume = {110},
  number = {22},
  pages = {220502},
  year = {2013},
  note = {\url{https://doi.org/10.1103/PhysRevLett.110.220502}},
  doi = {10.1103/PhysRevLett.110.220502}
}

@article{Cli22,
  title = {Coherent phase transfer for real-world twin-field quantum key distribution},
  author = {Clivati, Cecilia and Meda, Alice and Donadello, Simone and Virz{\`i}, Salvatore and Genovese, Marco and Levi, Filippo and Mura, Alberto and Pittaluga, Mirko and Yuan, Zhiliang and Shields, Andrew J. and Lucamarini, Marco and Degiovanni, Ivo Pietro and Calonico, Davide},
  journal = {Nat. Commun.},
  volume = {13},
  number = {1},
  pages = {157},
  year = {2022},
  doi = {10.1038/s41467-021-27808-1},
  note = {\url{https://doi.org/10.1038/s41467-021-27808-1}}
}

@ARTICLE{YAR02,
  author={Yariv, A.},
  journal={IEEE Photonics Technol. Lett.}, 
  title={Critical coupling and its control in optical waveguide-ring resonator systems}, 
  year={2002},
  volume={14},
  number={4},
  pages={483-485},
  doi={10.1109/68.992585},
  note = {\url{https://doi.org/10.1109/68.992585}},
  }

@article{Bravo07,
author = {Jorge Bravo-Abad and Alejandro Rodriguez and Peter Bermel and Steven G. Johnson and John D. Joannopoulos and Marin Solja\v{c}i\'{c}},
journal = {Opt. Express},
number = {24},
pages = {16161--16176},
publisher = {Optica Publishing Group},
title = {Enhanced nonlinear optics in photonic-crystal microcavities},
volume = {15},
month = {Nov},
year = {2007},
note = {\url{https://opg.optica.org/oe/abstract.cfm?URI=oe-15-24-16161}},
doi = {10.1364/OE.15.016161},
}

@article{Zhang23,
author = {Yi Zhang and Juniyali Nauriyal and Meiting Song and Marissa Granados Baez and Xiaotong He and Timothy Macdonald and Jaime Cardenas},
journal = {Opt. Mater. Express},
number = {1},
pages = {237--246},
publisher = {Optica Publishing Group},
title = {Engineered second-order nonlinearity in silicon nitride},
volume = {13},
month = {Jan},
year = {2023},
note = {\url{https://opg.optica.org/ome/abstract.cfm?URI=ome-13-1-237}},
doi = {10.1364/OME.478811},
}

@article{Ikeda08,
author = {Kazuhiro Ikeda and Robert E. Saperstein and Nikola Alic and Yeshaiahu Fainman},
journal = {Opt. Express},
number = {17},
pages = {12987--12994},
publisher = {Optica Publishing Group},
title = {Thermal and Kerr nonlinear properties of plasma-deposited silicon nitride/silicon dioxide waveguides},
volume = {16},
month = {Aug},
year = {2008},
note = {\url{https://opg.optica.org/oe/abstract.cfm?URI=oe-16-17-12987}},
doi = {10.1364/OE.16.012987},
}

@article{Carmon04,
author = {Tal Carmon and Lan Yang and Kerry J. Vahala},
journal = {Opt. Express},
number = {20},
pages = {4742--4750},
publisher = {Optica Publishing Group},
title = {Dynamical thermal behavior and thermal self-stability of microcavities},
volume = {12},
month = {Oct},
year = {2004},
note = {\url{https://opg.optica.org/oe/abstract.cfm?URI=oe-12-20-4742}},
doi = {10.1364/OPEX.12.004742},
}

@article{Glauber63,
  title = {The Quantum Theory of Optical Coherence},
  author = {Glauber, Roy J.},
  journal = {Phys. Rev.},
  volume = {130},
  issue = {6},
  pages = {2529--2539},
  numpages = {0},
  year = {1963},
  month = {Jun},
  publisher = {American Physical Society},
  doi = {10.1103/PhysRev.130.2529},
  note = {\url{https://link.aps.org/doi/10.1103/PhysRev.130.2529}},
}

@Article{Perez23,
author={Perez, Edgar F.
and Moille, Gr{\'e}gory
and Lu, Xiyuan
and Stone, Jordan
and Zhou, Feng
and Srinivasan, Kartik},
title={High-performance Kerr microresonator optical parametric oscillator on a silicon chip},
journal={Nat. Commun.},
year={2023},
month={Jan},
day={16},
volume={14},
number={1},
pages={242},
issn={2041-1723},
doi={10.1038/s41467-022-35746-9},
note = {\url{https://doi.org/10.1038/s41467-022-35746-9}},
}

@article{Garay13,
doi = {10.1088/1054-660X/23/1/015201},
note = {\url{https://dx.doi.org/10.1088/1054-660X/23/1/015201}},
year = {2012},
month = {nov},
publisher = {IOP Publishing},
volume = {23},
number = {1},
pages = {015201},
author = {Garay-Palmett, K and Jeronimo-Moreno, Y and U’Ren, A B},
title = {Theory of cavity-enhanced spontaneous four wave mixing},
journal = {Laser Phys.}
}

@article{Kipp18,
author = {Tobias J. Kippenberg  and Alexander L. Gaeta  and Michal Lipson  and Michael L. Gorodetsky },
title = {Dissipative Kerr solitons in optical microresonators},
journal = {Science},
volume = {361},
number = {6402},
pages = {eaan8083},
year = {2018},
doi = {10.1126/science.aan8083},
note = {\url{https://www.science.org/doi/abs/10.1126/science.aan8083}},
}

@article{Kipp04,
  title = {Kerr-Nonlinearity Optical Parametric Oscillation in an Ultrahigh-$Q$ Toroid Microcavity},
  author = {Kippenberg, T. J. and Spillane, S. M. and Vahala, K. J.},
  journal = {Phys. Rev. Lett.},
  volume = {93},
  issue = {8},
  pages = {083904},
  numpages = {4},
  year = {2004},
  month = {Aug},
  publisher = {American Physical Society},
  doi = {10.1103/PhysRevLett.93.083904},
  note = {\url{https://link.aps.org/doi/10.1103/PhysRevLett.93.083904}}
}

@article{Zhang22,
author = {Hao Zhang and Yifan Wu and Huashan Yang and Zongxin Ju and Zhe Kang and Jijun He and Shilong Pan},
journal = {Opt. Express},
number = {21},
pages = {37379--37393},
publisher = {Optica Publishing Group},
title = {Third-harmonic-assisted four-wave mixing in a chip-based microresonator frequency comb generation},
volume = {30},
month = {Oct},
year = {2022},
note = {\url{https://opg.optica.org/oe/abstract.cfm?URI=oe-30-21-37379}},
doi = {10.1364/OE.473472}
}

@article{ALM12,
title = {Four-wave mixing: Photon statistics and the impact on a co-propagating quantum signal},
journal = {Opt. Commun.},
volume = {285},
number = {12},
pages = {2956-2960},
year = {2012},
issn = {0030-4018},
doi = {https://doi.org/10.1016/j.optcom.2012.02.008},
note = {\url{https://www.sciencedirect.com/science/article/pii/S0030401812001423}},
author = {Álvaro J. Almeida and Nuno A. Silva and Paulo S. André and Armando N. Pinto}
}

@book{fox06,
  title={Quantum optics: an introduction},
  author={Fox, Anthony Mark},
  volume={15},
  year={2006},
  publisher={Oxford university press}
}

@book{Mand95, 
place={Cambridge}, 
title={Optical Coherence and Quantum Optics}, 
publisher={Cambridge University Press}, 
author={Mandel, Leonard and Wolf, Emil},
note = {\url{https://doi.org/10.1017/CBO9781139644105}},
doi = {https://doi.org/10.1017/CBO9781139644105},
year={1995}}

@book{Mig13,
title = {Single-Photon Generation and Detection},
author = {Alan Migdall and Sergey V. Polyakov and Jingyun Fan and Joshua C. Bienfang},
series = {Experimental Methods in the Physical Sciences},
publisher = {Academic Press},
volume = {45},
pages = {iii},
year = {2013},
issn = {1079-4042},
doi = {https://doi.org/10.1016/B978-0-12-387695-9.00017-2},
note = {\url{https://www.sciencedirect.com/science/article/pii/B9780123876959000172}}
}

@article{Che16,
  title = {Quantum dynamics of Kerr optical frequency combs below and above threshold: Spontaneous four-wave mixing, entanglement, and squeezed states of light},
  author = {Chembo, Yanne K.},
  journal = {Phys. Rev. A},
  volume = {93},
  issue = {3},
  pages = {033820},
  numpages = {24},
  year = {2016},
  month = {Mar},
  publisher = {American Physical Society},
  doi = {10.1103/PhysRevA.93.033820},
  note = {\url{https://link.aps.org/doi/10.1103/PhysRevA.93.033820}}
}

@article{Chembo_Modalexpansionapproach_2010,
  title = {Modal Expansion Approach to Optical-Frequency-Comb Generation with Monolithic Whispering-Gallery-Mode Resonators},
  author = {Chembo, Yanne K. and Yu, Nan},
  year = 2010,
  month = sep,
  journal = {Phys. Rev. A},
  volume = {82},
  number = {3},
  pages = {033801},
  publisher = {American Physical Society},
  doi = {10.1103/PhysRevA.82.033801},
  note = {\url{https://doi.org/10.1103/PhysRevA.82.033801}}
}

@article{Pasq18,
title = {Micro-combs: A novel generation of optical sources},
journal = {Phys. Rep.},
volume = {729},
pages = {1-81},
year = {2018},
note = {Micro-combs: A novel generation of optical sources},
issn = {0370-1573},
doi = {https://doi.org/10.1016/j.physrep.2017.08.004},
note = {\url{https://www.sciencedirect.com/science/article/pii/S0370157317303253}},
author = {Alessia Pasquazi and Marco Peccianti and Luca Razzari and David J. Moss and Stéphane Coen and Miro Erkintalo and Yanne K. Chembo and Tobias Hansson and Stefan Wabnitz and Pascal Del’Haye and Xiaoxiao Xue and Andrew M. Weiner and Roberto Morandotti},}

@article{Yariv00,
author = {A. Yariv},
title = {Universal relations for coupling of optical power between microresonators and dielectric waveguides},
journal = {Electron. Lett.},
volume = {36},
issue = {4},
pages = {321-322},
year = {2000},
doi = {10.1049/el:20000340},
note = {\url{https://digital-library.theiet.org/doi/abs/10.1049/el%3A20000340}},
}

@article{Chri:2011,
doi = {10.1088/1367-2630/13/3/033027},
note = {\url{https://doi.org/10.1088/1367-2630/13/3/033027}},
year = {2011},
month = {mar},
publisher = {},
volume = {13},
number = {3},
pages = {033027},
author = {Christ, Andreas and Laiho, Kaisa and Eckstein, Andreas and Cassemiro, Katiúscia N and Silberhorn, Christine},
title = {Probing multimode squeezing with correlation functions},
journal = {New J. Phys.},
}

@article{DelH:11,
  title = {Octave Spanning Tunable Frequency Comb from a Microresonator},
  author = {Del'Haye, P. and Herr, T. and Gavartin, E. and Gorodetsky, M. L. and Holzwarth, R. and Kippenberg, T. J.},
  journal = {Phys. Rev. Lett.},
  volume = {107},
  issue = {6},
  pages = {063901},
  numpages = {4},
  year = {2011},
  month = {Aug},
  publisher = {American Physical Society},
  doi = {10.1103/PhysRevLett.107.063901},
  note = {\url{https://link.aps.org/doi/10.1103/PhysRevLett.107.063901}},
}

@article{Mats:07,
author = {Andrey B. Matsko and Anatoliy A. Savchenkov and Nan Yu and Lute Maleki},
journal = {J. Opt. Soc. Am. B},
number = {6},
pages = {1324--1335},
publisher = {Optica Publishing Group},
title = {Whispering-gallery-mode resonators as frequency references. I. Fundamental limitations},
volume = {24},
month = {Jun},
year = {2007},
note = {\url{https://opg.optica.org/josab/abstract.cfm?URI=josab-24-6-1324}},
doi = {10.1364/JOSAB.24.001324},
}

@article{Cout23,
  title = {Applications of single photons to quantum communication and computing},
  author = {Couteau, Christophe and Barz, Stefanie and  Thomas, Durt and Thomas, Gerrits and Jan, Huwer and Robert, Prevedel and John, Rarity and Andrew, Shields and Gregor, Weihs},
  journal = {Nat. Rev. Phys.},
  volume = {5},
  number = {},
  pages = {326--338},
  year = {2023},
  doi = {10.1038/s42254-023-00583-2},
  note = {\url{https://doi.org/10.1038/s42254-023-00583-2}},
}

@article{Rari86,
  title = {Observation of sub-Poissonian light in parametric down-conversion},
  author = {Rarity, John G. and Tapster, Paul R. and Jakeman, Eric},
  journal = {Optics Commun.},
  volume = {62},
  number = {3},
  pages = {201--206},
  year = {1986},
  doi = {doi.org/10.1016/0030-4018(87)90028-9},
  note = {\url{https://doi.org/10.1016/0030-4018(87)90028-9}},
}

@article{Brid11,
author = {G. Brida and I. P. Degiovanni and M. Genovese and A. Migdall and F. Piacentini and S. V. Polyakov and I. Ruo Berchera},
journal = {Opt. Express},
number = {2},
pages = {1484--1492},
publisher = {Optica Publishing Group},
title = {Experimental realization of a low-noise heralded single-photon source},
volume = {19},
month = {Jan},
year = {2011},
note = {\url{https://opg.optica.org/oe/abstract.cfm?URI=oe-19-2-1484}},
doi = {10.1364/OE.19.001484},
}

@article{Poly09,
  title = {Quantum radiometry},
  author = {Polyakov, Sergey V. and Migdall, Alan L.},
  journal = {J. Mod. Opt.},
  volume = {56},
  number = {9},
  pages = {1045--1052},
  year = {2009},
  doi = {10.1080/09500340902919477},
  note = {\url{https://doi.org/10.1080/09500340902919477}},
}

@article{Thew07,
  title = {Quantum communication},
  author = {Thew, Rob and Gisin, Nicolas},
  journal = {Nat. Photonics},
  volume = {1},
  number = {},
  pages = {165--171},
  year = {2007},
  doi = {10.1038/nphoton.2007.22},
  note = {\url{https://doi.org/10.1038/nphoton.2007.22}},
}

@article{Obri09,
  title = {Photonic quantum technologies},
  author = {O'Brien, Jeremy L. and Furusawa, Akira and Vickovic, Jelena},
  journal = {Nat. Photonics},
  volume = {3},
  number = {},
  pages = {687--695},
  year = {2009},
  doi = {10.1038/nphoton.2009.229},
  note = {\url{https://doi.org/10.1038/nphoton.2009.229}},
}

@article{Flam19,
  title = {Photonic quantum information processing: a review},
  author = {Flamini, Fulvio and Spagnolo, Nicol{\`o} and Sciarrino, Fabio},
  journal = {Rep. Prog. Phys.},
  volume = {82},
  number = {1},
  pages = {016001},
  year = {2019},
  doi = {10.1088/1361-6633/aad5b2},
  note = {\url{https://doi.org/10.1088/1361-6633/aad5b2}},
}

@article{Mar03,
  title = {Long-distance teleportation of qubits at telecommunication wavelengths},
  author = {Marcikic, Ivan and de Riedmatten, Hugues and Tittel, Wolfgang and Zbinden, Hugo and Gisin, Nicolas},
  journal = {Nature},
  volume = {421},
  number = {},
  pages = {509--513},
  year = {2003},
  doi = {10.1038/nature01376},
   note = {\url{https://doi.org/10.1038/nature01376}},
}

@article{Asp82,
  title = {Experimental Test of Bell’s Inequalities Using Time-Varying Analyzers},
  author = {Aspect, Alain and Dalibard, Jean and Roger, G{\'e}rard},
  journal = {Phys. Rev. Lett.},
  volume = {49},
  number = {},
  pages = {1804--1807},
  year = {1982},
  doi = {10.1103/PhysRevLett.49.1804},
   note = {\url{https://doi.org/10.1103/PhysRevLett.49.1804}},
}

@article{Giu13,
  title = {Bell violation using entangled photons without the fair-sampling assumption},
  author = {Giustina, Marissa and Mech, Adrian and Ramelow, Sven and Wittmann, Bernhard and Kofler, Johannes and Beyer, J{\"o}rn and Lita, Adriana and Calkins, Brice and Gerrits, Thomas and Nam, Sae Woo and Ursin, Rupert and Zeilinger, Anton},
  journal = {Nature},
  volume = {497},
  number = {},
  pages = {227--230},
  year = {2013},
  doi = {10.1038/nature12012},
  note = {\url{https://doi.org/10.1038/nature12012}},
}

@article{Sha15,
  title = {Strong Loophole-Free Test of Local Realism},
  author = {Shalm, Lynden K. and
Meyer-Scott, Evan and
Christensen, Bradley G. and
Bierhorst, Peter and
Wayne, Michael A. and
Stevens, Martin J. and
Gerrits, Thomas and
Glancy, Scott and
Hamel, Deny R. and
Allman, Michael S. and
Coakley, Kevin J. and
Dyer, Shellee D. and
Hodge, Carson and
Lita, Adriana E. and
Verma, Varun B. and
Lambrocco, Camilla and
Tortorici, Edward and
Migdall, Alan L. and
Zhang, Yanbao and
Kumor, Daniel R. and
Farr, William H. and
Marsili, Francesco and
Shaw, Matthew D. and
Stern, Jeffrey A. and
Abell{\'a}n, Carlos and
Amaya, Waldimar and
Pruneri, Valerio and
Jennewein, Thomas and
Mitchell, Morgan W. and
Kwiat, Paul G. and
Bienfang, Joshua C. and
Mirin, Richard P. and
Knill, Emanuel and
Nam, Sae Woo},
  journal = {Phys. Rev. Lett.},
  volume = {115},
  number = {},
  pages = {250402},
  year = {2015},
  doi = {10.1103/PhysRevLett.115.250402},
  note = {\url{https://doi.org/10.1103/PhysRevLett.115.250402}},
}

@article{Wen19,
  title = {Entanglement distribution over a 96-km-long submarine optical fiber},
  author = {Wengerowsky, S{\"o}ren and
Joshi, Siddarth Koduru and
Steinlechner, Fabian and
Zichi, Julien R. and
Dobrovolskiy, Sergiy M. and
van der Molen, Ren{\'e} and
Losh, Johannes W. N. and
Zwillerg, Val and
Versteegh, Marijn A. M. and
Murai, Alberto and
Calonico, Davide and
Inguscio, Massimo and
H{\"u}bel, Hannes and
Bo, Liu and
Scheidl, Thomas and
Zeilinger, Anton and
Xuereb, Andre and
Ursin, Rupert},
  journal = {Proc. Natl. Acad. Sci.},
  volume = {116},
  number = {},
  pages = {6684--6688},
  year = {2019},
  doi = {10.1073/pnas.1818752116},
   note = {\url{https://doi.org/10.1073/pnas.1818752116}},
}

@article{Yin12,
  title = {Quantum teleportation and entanglement distribution over 100-kilometre free-space channels},
  author = {Yin, Juan and
Ren, Ji-Gang and
Lu, He and
Cao, Yuan and
Yong, Hai-Lin and
Wu, Yu-Ping and
Liu, Chang and
Liao, Sheng-Kai and
Zhou, Fei and
Jiang, Yan and
Cai, Xin-Dong and
Xu, Ping and
Pan, Ge-Sheng and
Jia, Jian-Jun and
Huang, Yong-Mei and
Yin, Hao and
Wang, Jian-Yu and
Chen, Yu-Ao and
Peng, Cheng-Zhi and
Pan, Jian-Wei},
  journal = {Nature},
  volume = {488},
  number = {},
  pages = {185--188},
  year = {2012},
  doi = {10.1038/nature11332},
  note = {\url{https://doi.org/10.1038/nature11332}},
}

@article{Jen02,
  title = {Entanglement swapping over 100 km optical fiber with independent entangled photon-pair sources},
  author = {Sun, Qi-Chao and
Jiang, Yang-Fan and
Mao, Ya-Li and
You, Li-Xing and
Zhang, Wei and
Zhang, Wei-Jun and
Jiang, Xiao and
Chen, Teng-Yun and
Li, Hao and
Huang, Yi-Dong and
Chen, Xian-Feng and
Wang, Zhen and
Fan, Jingyun and
Zhang, Qiang and
Pan, Jian-Wei},
  journal = {Optica},
  volume = {4},
  number = {},
  pages = {1214-1218},
  year = {2017},
  doi = {10.1364/OPTICA.4.001214},
  note = {\url{https://doi.org/10.1364/OPTICA.4.001214}},
}

@article{Tak02,
  title = {Quantum entanglement swapping with spontaneous parametric down-conversion},
  author = {Wang, Xiang-Bin and Shi, B. S. and Tomita, A. and Matsumoto, K.},
  journal = {Phys. Rev. A},
  volume = {69},
  number = {014303},
  pages = {},
  year = {2004},
  doi = {10.1103/PhysRevA.69.014303},
  note = {\url{https://doi.org/10.1103/PhysRevA.69.014303}},
}

@article{Matt96,
  title = {Dense Coding in Experimental Quantum Communication},
  author = {Mattle, Klaus and Weinfurter, Harald and Kwiat, Paul G. and Zeilinger, Anton},
  journal = {Phys. Rev. Lett.},
  volume = {76},
  number = {},
  pages = {4656--4659},
  year = {1996},
  doi = {10.1103/PhysRevLett.76.4656},
  note = {\url{https://doi.org/10.1103/PhysRevLett.76.4656}},
}

@article{Meda17,
doi = {10.1088/2040-8986/aa7b27},
note = {\url{https://doi.org/10.1088/2040-8986/aa7b27}},
year = {2017},
month = {aug},
publisher = {IOP Publishing},
volume = {19},
number = {9},
pages = {094002},
author = {Meda, A and Losero, E and Samantaray, N and Scafirimuto, F and Pradyumna, S and Avella, A and Ruo-Berchera, I and Genovese, M},
title = {Photon-number correlation for quantum enhanced imaging and sensing},
journal = {J. Opt.},
}

@article{Okam13,
  author    = {Togan, E. and Chu, Y. and Trifonov, A. S. and Jiang, L. and Maze, J. and Childress, L. and Dutt, M. V. G. and S{\o}rensen, A. S. and Hemmer, P. R. and Zibrov, A. S. and Lukin, M. D.},
  title     = {Quantum entanglement between an optical photon and a solid-state spin qubit},
  journal   = {Nature},
  year      = {2010},
  month     = {Aug},
  volume    = {466},
  number    = {7307},
  pages     = {730--734},
  doi       = {10.1038/nature09256},
  note = {\url{https://doi.org/10.1038/nature09256}},
  publisher = {Nature Publishing Group}
}

@book{Boyd08,
  title = {Nonlinear Optics},
  author = {Boyd, Robert W.},
  edition = {3},
  publisher = {Academic Press},
  address = {Amsterdam},
  year = {2008},
  isbn = {978-0-12-369470-6},
}

@article{Zapa23,
  title = {Advances in device-independent quantum key distribution},
  author = {Zapatero, Víctor and van Leent, Tim and Arnon-Friedman, Rotem and Liu, Wen- Zhao and Zhang, Qiang and Weinfurter, Harald and Curty, Marcos},
  journal = {npj Quantum Inf.},
  volume = {9},
  number = {10},
  pages = {},
  year = {2023},
  doi = {10.1038/s41534-023-00684-x},
  note = {\url{https://doi.org/10.1038/s41534-023-00684-x}},
}

@article{Ma18,
  title = {Simultaneous dual-band entangled photon pair generation using a silicon photonic microring resonator},
  author = {Ma, Chaoxuan and Mookherjea, Shayan},
  journal = {Quantum Sci. Technol.},
  volume = {3},
  number = {3},
  pages = {034001},
  year = {2018},
  doi = {10.1088/2058-9565/aab89a},
  note = {\url{https://doi.org/10.1088/2058-9565/aab89a}},
}

@article{Sam19,
author = {Farid Samara and Anthony Martin and Claire Autebert and Maxim Karpov and Tobias J. Kippenberg and Hugo Zbinden and Rob Thew},
journal = {Opt. Express},
number = {14},
pages = {19309--19318},
publisher = {Optica Publishing Group},
title = {High-rate photon pairs and sequential Time-Bin entanglement with Si3N4 microring resonators},
volume = {27},
month = {Jul},
year = {2019},
note = {\url{https://opg.optica.org/oe/abstract.cfm?URI=oe-27-14-19309}},
doi = {10.1364/OE.27.019309}
}

@article{Li25,
  author  = {Li, Bohan and Yuan, Zhiquan and Williams, James and Jin, Warren and Beckert, Adrian and Xie, Tian and Guo, Joel and Feshali, Avi and Paniccia, Mario and Faraon, Andrei and Bowers, John and Marandi, Alireza and Vahala, Kerry},
  title   = {Down-converted photon pairs in a high-Q silicon nitride microresonator},
  journal = {Nature},
  year    = {2025},
  volume  = {639},
  number  = {8056},
  pages   = {922--927},
  doi     = {10.1038/s41586-025-08662-3},
  note = {\url{https://doi.org/10.1038/s41586-025-08662-3}},
  issn    = {1476-4687}
}

@article{bonifacio1978optical,
  title={Optical bistability and cooperative effects in resonance fluorescence},
  author={Bonifacio, R and Lugiato, LA},
  journal={Phys. Rev. A},
  volume={18},
  number={3},
  pages={1129},
  year={1978},
  publisher={APS},
  note = {\url{https://doi.org/10.1103/PhysRevA.18.1129}}
}

@article{bonifacio1978photon,
  title={Photon statistics and spectrum of transmitted light in optical bistability},
  author={Bonifacio, R and Lugiato, LA},
  journal={Phys. Rev. Lett.},
  volume={40},
  number={15},
  pages={1023},
  year={1978},
  publisher={APS},
  note = {\url{https://doi.org/10.1103/PhysRevLett.40.1023}}
}

@article{bonifacio1976cooperative,
  title={Cooperative effects and bistability for resonance fluorescence},
  author={Bonifacio, R and Lugiato, LA},
  journal={Optics Commun.},
  volume={19},
  number={2},
  pages={172--176},
  year={1976},
  publisher={Elsevier},
  note = {\url{https://doi.org/10.1103/PhysRevA.18.1129}}
}

@article{Khi1934,
  author  = {Khintchine, A.},
  title   = {Korrelationstheorie der station{\"a}ren stochastischen Prozesse},
  journal = {Math. Ann.},
  year    = {1934},
  volume  = {109},
  number  = {1},
  pages   = {604--615},
  doi     = {10.1007/BF01449156},
  note = {\url{https://doi.org/10.1007/BF01449156}}
}

@article{Wie1930,
  author  = {Wiener, Norbert},
  title   = {Generalized harmonic analysis},
  journal = {Acta Math.},
  year    = {1930},
  volume  = {55},
  number  = {1},
  pages   = {117--258},
  doi     = {10.1007/BF02546511},
   note = {\url{https://doi.org/10.1007/BF02546511}}
}

@misc{OurSM,
  author = {},
  title  = {Supplemental Material},
  year   = {2026},
  note   = {See Supplemental Material at [URL/DOI] for additional material for this work.}
}

@article{Mat05,
  title = {Optical hyperparametric oscillations in a whispering-gallery-mode resonator: Threshold and phase diffusion},
  author = {Matsko, Andrey B. and Savchenkov, Anatoliy A. and Strekalov, Dmitry and Ilchenko, Vladimir S. and Maleki, Lute},
  journal = {Phys. Rev. A},
  volume = {71},
  issue = {3},
  pages = {033804},
  numpages = {10},
  year = {2005},
  month = {Mar},
  publisher = {American Physical Society},
  doi = {10.1103/PhysRevA.71.033804},
 note = {\url{https://link.aps.org/doi/10.1103/PhysRevA.71.033804}}
}

@manual{COMSOL64,
  title        = {{COMSOL Multiphysics{\textregistered} v. 6.4}},
  organization = {COMSOL AB},
  address      = {Stockholm, Sweden},
  year         = {2024},
   note = {\url{https://www.comsol.com}}
}

@article{Arb:13,
author = {Amir Arbabi and Lynford L. Goddard},
journal = {Opt. Lett.},
number = {19},
pages = {3878--3881},
publisher = {Optica Publishing Group},
title = {Measurements of the refractive indices and thermo-optic coefficients of Si3N4 and SiOx using microring resonances},
volume = {38},
month = {Oct},
year = {2013},
 note = {\url{https://opg.optica.org/ol/abstract.cfm?URI=ol-38-19-3878}},
doi = {10.1364/OL.38.003878},
}

@article{Tien:12,
author = {Chuen-Lin Tien and Tsai-Wei Lin},
journal = {Appl. Opt.},
number = {30},
pages = {7229--7235},
publisher = {Optica Publishing Group},
title = {Thermal expansion coefficient and thermomechanical properties of SiNx thin films prepared by plasma-enhanced chemical vapor deposition},
volume = {51},
month = {Oct},
year = {2012},
 note = {\url{https://opg.optica.org/ao/abstract.cfm?URI=ao-51-30-7229}},
doi = {10.1364/AO.51.007229},
}

@misc{Kampen81,
  title={van Kampen: Stochastic Processes in Physics and Chemistry},
  author={Quack, MNG},
  year={1981},
  publisher={north holland publishing company, amsterdam}
}

@book{sieg43,
  title={On the fluctuations in signals returned by many independently moving scatterers},
  author={Siegert, AJF},
  year={1943},
  publisher={Radiation Laboratory, Massachusetts Institute of Technology}
}

@Article{chen24lin,
AUTHOR = {Chen, Jia-Qi and Chen, Chao and Sun, Jing-Jing and Zhang, Jian-Wei and Liu, Zhao-Hui and Qin, Li and Ning, Yong-Qiang and Wang, Li-Jun},
TITLE = {Linewidth Measurement of a Narrow-Linewidth Laser: Principles, Methods, and Systems},
JOURNAL = {Sensors},
VOLUME = {24},
YEAR = {2024},
NUMBER = {11},
ARTICLE-NUMBER = {3656},
note = {\url{https://www.mdpi.com/1424-8220/24/11/3656}},
PubMedID = {38894446},
ISSN = {1424-8220},
DOI = {10.3390/s24113656}
}

@article{Ji25,
  author  = {Ji, Xinru and Li, Xurong and Qiu, Zheru and Wang, Rui Ning and Divall, Marta and Gelash, Andrey and Lihachev, Grigory and Kippenberg, Tobias J.},
  title   = {Deterministic soliton microcombs in {Cu}-free photonic integrated circuits},
  journal = {Nature},
  year    = {2025},
  volume  = {646},
  number  = {8086},
  pages   = {843--849},
  month   = {oct},
  doi     = {10.1038/s41586-025-09598-4},
   note = {\url{https://doi.org/10.1038/s41586-025-09598-4}}
}

@article{PerLo:21,
author = {Daniel P\'{e}rez-L\'{o}pez and Ana Guti\'{e}rrez and Jos\'{e} Capmany},
journal = {Opt. Express},
number = {6},
pages = {9043--9059},
publisher = {Optica Publishing Group},
title = {Silicon nitride programmable photonic processor with folded heaters},
volume = {29},
month = {Mar},
year = {2021},
 note = {\url{https://opg.optica.org/oe/abstract.cfm?URI=oe-29-6-9043}},
doi = {10.1364/OE.416053},
}

@article{Josh16,
author = {Chaitanya Joshi and Jae K. Jang and Kevin Luke and Xingchen Ji and Steven A. Miller and Alexander Klenner and Yoshitomo Okawachi and Michal Lipson and Alexander L. Gaeta},
journal = {Opt. Lett.},
number = {11},
pages = {2565--2568},
publisher = {Optica Publishing Group},
title = {Thermally controlled comb generation and soliton modelocking in microresonators},
volume = {41},
month = {Jun},
year = {2016},
 note = {\url{https://opg.optica.org/ol/abstract.cfm?URI=ol-41-11-2565}},
doi = {10.1364/OL.41.002565},
}

@article{Herr26,
  title={Frequency combs and coherent dissipative structures in nonlinear optical microresonators},
  author={Herr, Tobias and Tikan, Alexey and Kippenberg, Tobias J},
  journal={arXiv:2604.05897},
  year          = {2026},
  eprint        = {2604.05897},
  primaryClass  = {physics.optics},
  doi           = {10.48550/arXiv.2604.05897},
   note = {\url{https://arxiv.org/abs/2604.05897}}
}

@article{LLe87,
  title = {Spatial Dissipative Structures in Passive Optical Systems},
  author = {Lugiato, L. A. and Lefever, R.},
  journal = {Phys. Rev. Lett.},
  volume = {58},
  issue = {21},
  pages = {2209--2211},
  numpages = {0},
  year = {1987},
  month = {May},
  publisher = {American Physical Society},
  doi = {10.1103/PhysRevLett.58.2209},
   note = {\url{https://link.aps.org/doi/10.1103/PhysRevLett.58.2209}}
}

@InCollection{Genovese1,
	author       =	{Myrvold, Wayne and Genovese, Marco and Shimony, Abner},
	title        =	{{Bell’s Theorem}},
	booktitle    =	{The {Stanford} Encyclopedia of Philosophy},
	editor       =	{Edward N. Zalta and Uri Nodelman},
	 note = {\url{https://plato.stanford.edu/archives/spr2024/entries/bell-theorem/}},
	year         =	{2024},
	edition      =	{{S}pring 2024},
	publisher    =	{Metaphysics Research Lab, Stanford University}
}

@article{Genovese2,
  author   = {Brida, G. and Genovese, M. and Ruo Berchera, I.},
  title    = {Experimental realization of sub-shot-noise quantum imaging},
  journal  = {Nat. Photonics},
  year     = {2010},
  volume   = {4},
  number   = {4},
  pages    = {227--230},
  doi      = {10.1038/nphoton.2010.29},
   note = {\url{https://doi.org/10.1038/nphoton.2010.29}}
}

@article{Levy2010,
  author   = {Levy, Jacob S. and Gondarenko, Alexander and Foster, Mark A. and Turner-Foster, Amy C. and Gaeta, Alexander L. and Lipson, Michal},
  title    = {{CMOS}-compatible multiple-wavelength oscillator for on-chip optical interconnects},
  journal  = {Nat. Photonics},
  year     = {2010},
  volume   = {4},
  number   = {1},
  pages    = {37--40},
  doi      = {10.1038/nphoton.2009.259},
   note = {\url{https://doi.org/10.1038/nphoton.2009.259}}
}

@article{Kras19,
  author  = {Krasnokutska, Inna and Tambasco, Jean-Luc J. and Peruzzo, Alberto},
  title   = {Tunable large free spectral range microring resonators in lithium niobate on insulator},
  journal = {Scientific Reports},
  year    = {2019},
  volume  = {9},
  number  = {1},
  pages   = {11086},
  doi     = {10.1038/s41598-019-47231-3},
  note = {\url{https://doi.org/10.1038/s41598-019-47231-3}}
}

@article{Erk:14,
author = {Miro Erkintalo and St\'{e}phane Coen},
journal = {Opt. Lett.},
number = {2},
pages = {283--286},
publisher = {Optica Publishing Group},
title = {Coherence properties of Kerr frequency combs},
volume = {39},
month = {Jan},
year = {2014},
note = {\url{https://opg.optica.org/ol/abstract.cfm?URI=ol-39-2-283}},
doi = {10.1364/OL.39.000283},
}

@article{Yan25,
  author  = {Yan, Wenhan and Zheng, Xiaodong and Wen, Wenjun and Lu, Liangliang and Du, Yifeng and Lu, Yan-Qing and Zhu, Shining and Ma, Xiao-Song},
  title   = {A measurement-device-independent quantum key distribution network using optical frequency comb},
  journal = {npj Quantum Inf.},
  year    = {2025},
  volume  = {11},
  number  = {1},
  pages   = {97},
  doi     = {10.1038/s41534-025-01052-7},
  note = {\url{https://doi.org/10.1038/s41534-025-01052-7}},
  issn    = {2056-6387}
}

@article{BROWN56,
  author  = {Hanbury Brown, R. and Twiss, R. Q.},
  title   = {Correlation between Photons in two Coherent Beams of Light},
  journal = {Nature},
  year    = {1956},
  volume  = {177},
  number  = {4497},
  pages   = {27--29},
  doi     = {10.1038/177027a0},
  note = {\url{https://doi.org/10.1038/177027a0}},
  issn    = {1476-4687}
}

@article{Vir24,
doi = {10.1088/2058-9565/ad6a37},
year = {2024},
month = {aug},
publisher = {IOP Publishing},
volume = {9},
number = {4},
pages = {045027},
author = {Virzì, Salvatore and Rebufello, Enrico and Atzori, Francesco and Avella, Alessio and Piacentini, Fabrizio and Lussana, Rudi and Cusini, Iris and Madonini, Francesca and Villa, Federica and Gramegna, Marco and Cohen, Eliahu and Pietro Degiovanni, Ivo and Genovese, Marco},
title = {Entanglement-preserving measurement of the Bell parameter on a single entangled pair},
journal = {Quantum Science and Technology},
note    = {\url{https://doi.org/10.1088/2058-9565/ad6a37}}
}

@article{Genovese3,
  title={Experimental quantum enhanced optical interferometry},
  author={Genovese, Marco},
  journal={AVS Quantum Science},
  volume={3},
  number={4},
  year={2021},
  publisher={AIP Publishing},
  doi = {10.1116/5.0062114},
  note    = {\url{https://doi.org/10.1116/5.0062114}}
}

@article{Ong26,
author = {Ong, Kenny Y. K. and Chia, Xavier X. and Aadhi, A. and Chen, George F. R. and Choi, Ju Won and Sohn, Byoung Uk and Chowdury, Amdad and Tan, Dawn T. H.},
title = {Deterministic, Reconfigurable Micro-Resonator Soliton Crystals for Intensity-Modulated Direct Detection Data Transmission},
journal = {Laser Photonics Rev.},
volume = {20},
number = {4},
pages = {e00974},
doi = {https://doi.org/10.1002/lpor.202500974},
note    = {\url{https://onlinelibrary.wiley.com/doi/abs/10.1002/lpor.202500974}},
year = {2026}
}

@article{Mah23,
  author  = {Mahmudlu, Hatam and Johanning, Robert and van Rees, Albert and Khodadad Kashi, Anahita and Epping, J{\"o}rn P. and Haldar, Raktim and Boller, Klaus-J. and Kues, Michael},
  title   = {Fully on-chip photonic turnkey quantum source for entangled qubit/qudit state generation},
  journal = {Nat. Photonics},
  year    = {2023},
  volume  = {17},
  number  = {6},
  pages   = {518--524},
  doi     = {10.1038/s41566-023-01193-1},
  note    = {\url{https://doi.org/10.1038/s41566-023-01193-1}},
  issn    = {1749-4893},
}

\newpage
\renewcommand{\theequation}{S.\arabic{equation}}
\setcounter{equation}{0}
\renewcommand{\thefigure}{S\arabic{figure}}
\setcounter{figure}{0}
\renewcommand{\thesection}{S\arabic{section}}
\setcounter{section}{0}

\title{Spectral evolution of two-photon emission in microresonators: Supplemental Document}

\begin{figure}[hptb]
\includegraphics[width=1\linewidth, keepaspectratio]{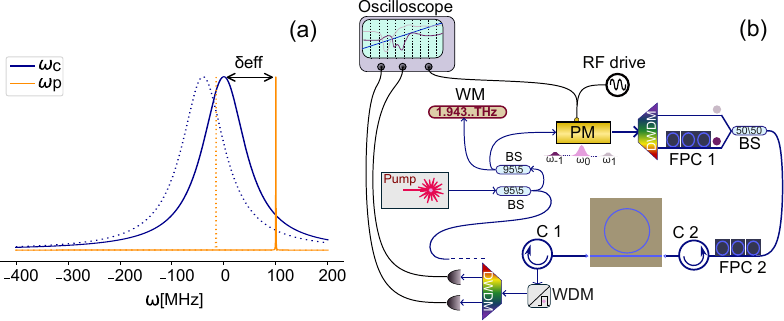}
\caption{\label{fig:det_eff} (a) Effective detuning for two values of $\omega_p$ (solid and dotted lines). The solid orange narrow feature represents the pump laser, while the solid blue broad one corresponds to the microresonator resonance. The effective detuning, indicated by the black arrow, is defined as the frequency difference between the pump laser and the instantaneous cavity resonance. In the next step (dotted curves), $\omega_p$ is decreased by \qty{110}{MHz}, which would nominally shift $\delta_\text{eff}$ from blue- to red-detuned. However, the increased circulating power inside the resonator induces thermal effects that shift the cavity resonance as well, maintaining a blue-detuned configuration. (b) Scheme for the measurement of $\delta_\text{eff}$. BS Beam Splitter, PM Phase Modulator, RF Radio Frequency, FPC Fiber Polarization Controller, DWDM Dense Wavelength Division Multiplexer, PD PhotoDiode, C circulator, WM WaveMeter.}
\end{figure}

\setcounter{section}{0}
\renewcommand*{\theHsection}{chY.\the\value{section}}

\section{Effective detuning measurement \label{SMSec:eff_det}}
The effective detuning $\delta_{\text{eff}}$ quantifies the instantaneous frequency offset between the pump laser $\omega_p$ and the shifted cavity resonance $\omega_c$ at each value of $\omega_p$, see Fig.~\ref{fig:det_eff}(a). As the pump frequency approaches the cavity resonance, the circulating optical power inside the microresonator increases, which heats the microring inducing thermal expansion and thermo-optic effects. These modify the effective optical path (OP) of the cavity resulting in a power-dependent redshift of the cavity resonance frequency $\omega_0$~\cite{Carmon04}. Simultaneously, nonlinear effects also shift the resonance.
Consequently, the detuning $\delta_{\text{eff}}$ does not solely reflect the externally imposed change in $\omega_p$, but also incorporates the cavity's thermally and nonlinearly induced response.\\
\begin{figure}
\includegraphics[width=1\linewidth, keepaspectratio]{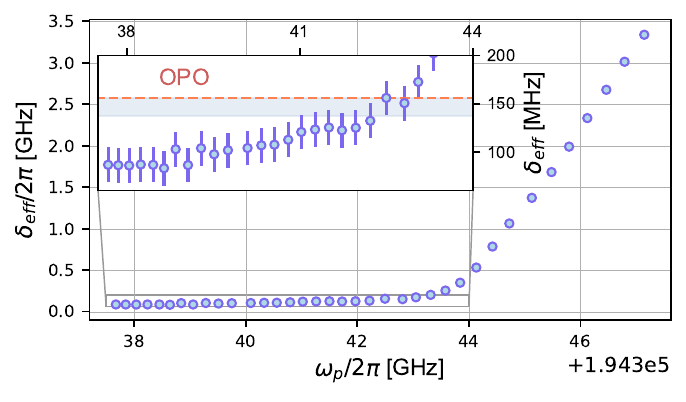}
\caption{\label{fig:det_eff_res} Effective detuning measurements as a function of $\omega_p$. Inset, zoom for $\delta_{\text{eff}} \sim \gamma$. The dashed line corresponds to $\delta_{\text{eff}} \sim 1.25\gamma$. The input power set by the EDFA, see Fig.~\ref{fig:Sec_exp} in Main text, is \qty{200}{mW}, the RF scan is [34, 36.5]$\,$GHz.}
\end{figure}
To characterize the effective detuning $\delta_{\text{eff}}(\omega_p)$ during a measurement, we monitor the shift of the cavity resonance while scanning the pump frequency $\omega_p$. Direct monitoring of the pumped resonator mode is experimentally challenging due to detector saturation and high background noise caused by the high DC pump power.  Instead, we employ a weak scanning probe on an adjacent sideband mode, which enables a low-noise, high-sensitivity tracking of the cavity resonance shift.\\
The experimental scheme is illustrated in Fig.~\ref{fig:det_eff}(b). The laser beam is split into two paths: 5$\%$ of the power is amplified using an erbium-doped fiber amplifier (EDFA) and used as the pump for entangled photon-pair generation. The remaining 95$\%$ is split again. 5$\%$ is sent to the wavemeter for monitoring of $\omega_p$, while 95$\%$ is routed through an optical phase modulator. The modulation frequency is chosen such that the fourth-order sidebands probe the cavity resonances located one free spectral range (FSR) to the left and to the right of the main cavity mode used for generation. After the modulator, these sidebands are selected by DWDM filters centered at ITU channels 40.5 (center \qty{1544.7}{nm}) and 42.5 (center \qty{1542.5}{nm}). The RF frequency is continuously swept to estimate the points $\omega_\mathrm{\pm 1}$ of minimum transmission of the resonance feature. $\delta_{\text{eff}}(\omega_p)=\omega_p-(\omega_{+1}+\omega_{-1})/2$ is then obtained as the half-difference of these minima. The typical modulation range is [34, 36.5]$\,$GHz while performing the measurement presented in this work, i.e. for $\delta_{\text{eff}}(\omega_p)$ of the order of $1.2-1.5\,\gamma$. 

We also studied the behavior of $\delta_{\text{eff}}(\omega_p)$ at larger detunings. Far from the cold cavity resonance, $\delta_{\text{eff}} \gg \gamma$, the circulating power is small and the cavity mode remains essentially unchanged $\omega_c \rightarrow\omega_0$. In this regime, $\delta_{\text{eff}}(\omega_p) \sim \omega_p-\omega_0$. This behavior is visible in Fig.~\ref{fig:det_eff_res} as the linear trend on the right. 

In contrast, when $\delta_{\text{eff}} \sim \gamma$, the circulating power increases and the thermally and nonlinearly induced red shift of the cavity resonance becomes significant. This reduces the rate of change of $\delta_{\text{eff}}$ with respect to $\omega_p$. This behavior is illustrated in the inset of the figure. 
For the input power used, the regime shown in the inset corresponds to operating in the optical parametric oscillation regime, and the threshold to enter this regime is indicated by the red star in figure.

\section{Self-interference measurement \label{SMSec:int_meas}}

\begin{figure}
\includegraphics[width=1\linewidth, keepaspectratio]{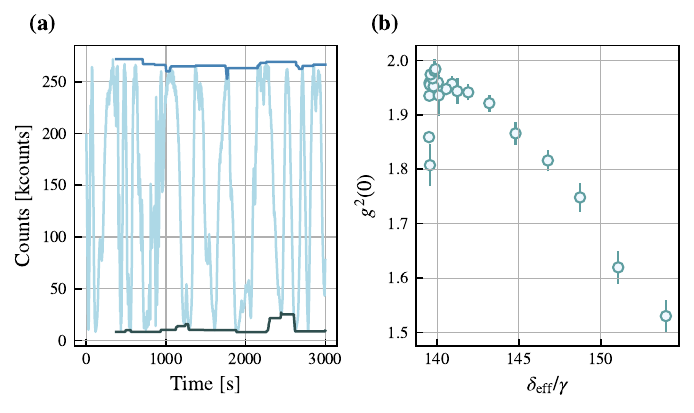}
\caption{(a) Self-interference fringes for idler mode. The measurement is performed for an effective detuning $\delta_\text{eff} = \qty{139(1)}{MHz}$ and path difference $L =\qty{0}{m}$. The single acquisition duration is \qty{1}{s}. This graph corresponds to the first point in Fig. 2 in Main text. 
(b) Second-order coherence function $g^{(2)}(t_\text{delay})$ evaluated at $t_\text{delay} = 0$ as a function of the parameter $\delta_\text{eff}/\gamma$.
}\label{fig:SM_int}
\end{figure}

As described in the main text, we measure the first-order correlation function $|g^{(1)}(t)|$ of a single mode of the photon pair generated by the FWM process through a self-interference experiment. The selected mode is split into two arms, which are then recombined after introducing a relative delay. This delay is implemented with an optical fiber of length $L$; since all other components in the two paths are identical, the delay is given by $t = L n_\text{f}/c$, where $c/n_\text{f}$ is the velocity of light in a fiber with refractive index $n_\text{f}$. After recombination, interference occurs only if the delay is shorter than the photon coherence length $\Delta L$, see Eq.~(5) in Main text.
For $L =$ \qty{0}{m} the expected maximum interference fringe visibility is $V_0 = 1$ (see Eq.~(4) in Main text).  
In practice, however, the measured value of $V_0$ is reduced by several factors~\cite{Mand95} such as (a) an imbalance between the arms at the recombination stage, (b) a polarization mismatch (the electric fields must overlap in polarization to interfere), and (c) an imperfect spatial mode matching. We note that, beyond lowering the maximum visibility $V_0$, these factors affect the measured visibility $V$ at each delay time. For this reason, they must be re-optimized whenever a new fiber segment is inserted. To minimize (a), all components in the two arms are kept identical, except for the delay fiber, see Fig. 2(e) in Main text. 
In addition to introducing a relative delay, the extra fiber also introduces absorption loss. For our standard single-mode telecom fiber this is of the order of \qty{0.2}{dB/km}. For coherence lengths of the order of $\sim$ m ($\tau \sim$ \qty{10}{ns}) this contribution is negligible.
Minimization of effect (b) is actively performed for each fiber length $L$ using two polarization controllers. To this end, only one arm at a time (e.g. arm 1 while arm 2 is disconnected, and vice versa) is sent to the SNSPD channel, and the counts are maximized by adjusting the corresponding paddle. Since the detection efficiency of the SNSPDs is polarization dependent~\cite{Mig13}, this procedure ensures that each arm is individually optimized. As a result, when both arms are combined, their polarizations are effectively matched.
Finally, as the setup relies exclusively on fiber-integrated components we assume (c) to be inherently optimized.
Fig.~\ref{fig:SM_int} (a) shows an example of the interference fringes, in particular for the case $L = \qty{0}{m}$. From this measurement we obtain a maximum fringe visibility of 0.92.

\section{Second-order correlation measurements \label{SMSec:g2}}

Autocorrelation functions and coincidence histograms are acquired using a time-tagger operating in bidirectional histogram mode. The time-tagger records the relative delay between detection events on two SNSPD channels. One channel is assigned as the start event, registering events at $t_\text{start}$ and the other as the stop event, registering at $t_\text{stop}$. Any fixed optical path difference between the fibers feeding the two channels introduces a constant delay $t_\text{op}$ shifting the peak of the coincidence histogram from $t_\text{delay} = 0$ to $t_\text{delay} = t_\text{op}$. Although this offset can be removed by postprocessing, we avoid it altogether by using a common optical fiber for both modes and splitting them immediately before the SNSPDs, via a DWDM (for cross-correlations) or a 50/50 beam splitter (for autocorrelations). The small offset remaining does not affect the measurement.

The temporal accuracy of each timestamp is limited by the SNSPD timing jitter, of the order of \qty{100}{ps}, which sets the minimum histogram bin size (detection window). This jitter is more than an order of magnitude shorter than the characteristic timescale imposed by the cavity decay (around \qty{3}{ns}), and so of the fastest decay features observed in our correlation functions. The experiment therefore operates in a genuinely time-resolved regime. In this regime, the measured autocorrelation histogram is equivalent to the second-order correlation function $g^{(2)}(t_\text{delay})$, after subtraction of the background arising at long delay times from uncorrelated noise events~\cite{Chri:2011}.

While an ideal coherent source yields $g^{(2)}(0) = 1$, a thermal or bunched source is expected to exhibit 
$g^{(2)}(0) = 2$. In practice, however, accidental coincidences and background events reduce the measured value below its ideal limit. Fig.~\ref{fig:SM_int} (b) illustrates this behavior. As we move towards resonance, the idler output power increases improving the signal-to-noise ratio. The measured $g^{(2)}(0)$ then approaches the expected thermal value of 2. The last two points are associated with the onset of coherent emission from the system, where $g^{(2)}(0)$ transitions abruptly toward unity.

\section{Heterodyne measurement \label{SMSec:Het}}

For the heterodyne measurement, we employ a reference local oscillator (LO) derived from a frequency comb ($f_\text{rep} = 250$ MHz) phase-locked to an ultrastable cavity, which ensures a tooth linewidth of $< 10$ Hz. We use a DWDM channel to select the specific comb tooth closest to the idler mode. A single tooth provides approximately $20\text{ }\mu\text{W}$ of power; when combined with the microring-generated mode via a 50/50 splitter, the power contributing to the beat note is roughly $10\text{ }\mu\text{W}$. To maximize the measurement range, the microring-generated mode undergoes one or two stages of optical amplification before combination. The resulting beat signal is further amplified and passed through a 120 MHz low-pass filter; this isolates the primary heterodyne beat and suppresses neighboring teeth (spaced by 250 MHz) that fall within the DWDM optical bandwidth. The filtered signal is then recorded using an RF spectrum analyzer. For each data point, we acquire six independent measurements (representative trace in Fig. 2(h) in Main text) while manually recording the effective detuning ($\delta_{\text{eff}}$) in parallel. The data presented in Fig. 4 in Main text represent the average of these six measurements, that are consequently fit by a lorenztian profile.

\section{Theory \label{SMSec:theory}}
We give some details about the derivation of main formulae in subsection~\ref{Sec:Theory}.
The linearized equation of motion \eqref{eq:mat_stability_sidebands} for the mean values of the sidebands' quantum fluctuations $(\langle \hat{a}_s(t)\rangle, \langle \hat{a}_i^\dagger(t)\rangle)$ 
has the exact solution
\begin{equation}
\begin{pmatrix}
 \langle \hat{a}_s(t)\rangle\\
				 \langle \hat{a}_i^\dagger(t)\rangle
\end{pmatrix} = e^{-\frac{\gamma}{2} t}\begin{pmatrix}
				 \cosh(t\bar{\rho})+i\frac{\delta_s+2{g} {n_0}}{\bar{\rho}} \sinh(t\bar{\rho})& i\frac{ {g} \alpha_0^2}{\bar{\rho}}\sinh(t\bar{\rho})\\
				 -i \frac{{g} (\alpha_0^\ast)^2}{\bar{\rho}} \sinh(t\bar{\rho})&\cosh(t\bar{\rho})-i\frac{\delta_s+2{g} {n_0}}{\bar{\rho}} \sinh(t\bar{\rho})
				 \end{pmatrix} 
\begin{pmatrix}
 \langle \hat{a}_s(0)\rangle\\
				 \langle \hat{a}_i^\dagger(0)\rangle
\end{pmatrix} \,,
\label{eq:exactsol}
\end{equation} 

For the two Bogoliubov operators 
\begin{equation}
 \begin{pmatrix}
 \hat{b}\\
 \hat{c}^\dagger
 \end{pmatrix} = 
 \begin{pmatrix}
 \cosh\theta e^{-i\frac{\phi+\psi}{2}} & \sinh\theta e^{i\frac{\psi-\phi}{2}}\\
 \sinh\theta e^{-i\frac{\psi-\phi}{2}} & \cosh\theta e^{i\frac{\phi+\psi}{2}}
\end{pmatrix}
\begin{pmatrix}
 \hat{a}_s\\
	\hat{a}_i^\dagger
\end{pmatrix}\,,
\end{equation}
where the parameters $\theta$, $\phi$ are fixed by the condition
\begin{equation}
 (\delta_s+2{g} {n_0}) \sinh 2\theta +{g} \alpha_0^2 (\cosh^2\theta e^{i\phi} +\sinh^2\theta e^{-i\phi}) =0 \,,
\end{equation}
namely
\begin{align}
 2(\delta_s+2{g} {n_0}) \sinh 2\theta +g(\alpha_0^2 +(\alpha_0^\ast)^2)\cosh2\theta \cos\phi +i g(\alpha_0^2-(\alpha_0^\ast)^2) \sin\phi=0\,,\\
 (\alpha_0^2-(\alpha_0^\ast)^2)\cosh2\theta \cos\phi +i (\alpha_0^2 +(\alpha_0^\ast)^2)\sin\phi=0\,,
\end{align}
we then have
\begin{align}
 \langle \hat{b}(t)\rangle = & e^{-\frac{\gamma}{2} t}(\cosh(t\bar{\rho})+ \sinh(t\bar{\rho}))\langle \hat{b}(0)\rangle\,,\\
 \langle \hat{c}^\dagger(t)\rangle = & e^{-\frac{\gamma}{2} t}(\cosh(t\bar{\rho})-\sinh(t\bar{\rho}))\langle \hat{c}^\dagger(0)\rangle\,.
\end{align}
We note that, for imaginary $\bar{\rho}= i\Omega$, 
\begin{equation}
 \langle \hat{b}(t)\rangle =  e^{-\frac{\gamma}{2} t+ i t \Omega}\langle \hat{b}(0)\rangle\,,\qquad
 \langle \hat{c}^\dagger(t)\rangle =  e^{-\frac{\gamma}{2} t -i t \Omega}\langle \hat{c}^\dagger(0)\rangle\,.
\end{equation}
When $\bar{\rho}$ is real, we have instead
\begin{equation}
 \langle \hat{b}(t)\rangle =  e^{-(\frac{\gamma}{2}-\bar{\rho}) t}\langle \hat{b}(0)\rangle\,,\qquad
 \langle \hat{c}^\dagger(t)\rangle =  e^{-(\frac{\gamma}{2}+\bar{\rho})t}\langle \hat{c}^\dagger(0)\rangle\,.
\end{equation}
The stability condition \eqref{eq:stability_sidebands} is such that the absolute values of both amplitudes are decreasing with time. 

Equation $C(n)=0$ has either one or two solutions that are dynamically stable to small perturbations~\cite{bonifacio1976cooperative,bonifacio1978optical,bonifacio1978photon}.
In particular, for driving amplitudes lower than the critical pump amplitude $|\beta|\le \beta_c = \gamma^{3/2}/(3^{3/2}g)^{1/2}$ only one real stable solution for $n$ exists for any $\delta_0$. In the supercritical regime, there is an interval of detunings for which the mean-field equations of motion can support two dynamically stable steady states. However, if the system is initialized in one of the mean-field steady states, it will explore the phase space in the vicinity of that solution on the natural time scales of the dynamics. 

The linearization approach allows for an analytic solution of the relevant modes' spectra. In order to evaluate the sideband first-order correlation function $g^{(1)}$, two operators are needed: the total sideband occupation $\hat{K} =\frac12 (\hat{a}_s^\dagger\hat{a}_s +\hat{a}_i^\dagger\hat{a}_i +1)$, and the pair operator $\hat{K}_+ = \hat{a}_s^\dagger\hat{a}_i^\dagger$, with its Hermitian conjugate $\hat{K}_- = \hat{a}_s\hat{a}_i$. These form a $SU(1,1)$ algebra typical of models with two bosonic fields~\cite{Chiribella_ApplicationsgroupSU1_2006}, that eases the analytic derivation of their evolution from the Lindblad~\eqref{eq:Lindblad} as~\cite{GiaccariOPO2026}
\begin{equation}
\label{eq:su11generatorsEoM}
 \frac{d}{dt} \begin{pmatrix}
				 \langle \hat{K}_+(t)\rangle\\
				 \langle \hat{K}(t)\rangle\\
    \langle \hat{K}_-(t)\rangle 
				 \end{pmatrix} = \begin{pmatrix}
     -\gamma -2i(\delta_s+2g n_0) & -2i g (\alpha_0^\ast)^2 & 0\\
     i g \alpha_0^2 & -\gamma & -i g (\alpha_0^\ast)^2\\
     0 & 2i g \alpha_0^2 & -\gamma +2i(\delta_s+2g n_0) 
     \end{pmatrix}\begin{pmatrix}
				 \langle \hat{K}_+(t)\rangle\\
				 \langle \hat{K}(t)\rangle\\
    \langle \hat{K}_-(t)\rangle 
				 \end{pmatrix}+ \begin{pmatrix}
      0\\ \frac{\gamma}{2} \\0
     \end{pmatrix}\,,
\end{equation}
We can use \eqref{eq:exactsol} to find the homogeneous solutions to \eqref{eq:su11generatorsEoM}. In particular, given the initial conditions
\begin{equation}
 \begin{pmatrix}
				 \langle \hat{K}_+(0)\rangle\\
				 \langle \hat{K}(0)\rangle\\
    \langle \hat{K}_-(0)\rangle 
				 \end{pmatrix} = 
     \begin{pmatrix}
      0 \\ \frac{1}{2}\\ 0
     \end{pmatrix}\,,
\end{equation}
we have
\begin{align}
 \langle \hat{K}(t)\rangle = &-\frac{{g}^2{n_0}^2}{2\Delta}e^{-{\gamma} t}\left(|\cosh(t\bar{\rho})|^2 + \frac{g^{2}n_0^2 + (\delta_s+2g n_0)^2}{|\bar{\rho}|^2}|\sinh(t\bar{\rho})|^2 \right)+\frac{\gamma^2 + 4(\delta_s+2g n_0)^2}{8\Delta}\,.\\
\end{align}
We define the unnormalized first-order correlation function as 
\begin{equation}
 G^{(1)}(\tau) =\lim_{t\to\infty}\frac{1}{2} \langle\hat{a}_s^\dagger(t)\hat{a}_s(t+\tau) +\hat{a}_i^\dagger(t)\hat{a}_i(t+\tau)\rangle\,,
\end{equation}
and the normalized one as $g^{(1)}(\tau)=G^{(1)}(\tau)/G^{(1)}(0)$.

By the quantum regression theorem we have
\begin{align}
 G^{(1)}(\tau) = & e^{-\frac{\gamma}{2} \tau}\left(\Re\left[\cosh(\tau\bar{\rho})\right]+i(\delta_s+2g n_0)\Re\left[\frac{\sinh(\tau\bar{\rho})}{\bar{\rho}}\right] \right)\left(\langle\hat{K}\rangle-\frac12\right) \nonumber\\
 \phantom{=}& +i g \alpha_0^2 e^{-\frac{\gamma}{2} \tau} \Re\left[\frac{\sinh(\tau\bar{\rho})}{\bar{\rho}}\right] \langle \hat{K}_+\rangle\nonumber\\
 = & \frac{g^{2} n_0^2}{2\Delta}e^{-\frac{\gamma}{2} \tau}\left(\Re\left[\cosh(\tau\bar{\rho})\right] + \frac{\gamma}{2}\Re\left[\frac{\sinh(\tau\bar{\rho})}{\bar{\rho}}\right]\right)\,.
 \label{Eq:g1_general}
\end{align}
These analytic formulae, obtained from the large-occupation and mean-field approximation for the pumped mode, whereby SPM and XPM interactions involving only sidebands are negligible, can be compared with the simulations performed using the Julia library QuantumCumulants, which provides an implementation of cumulant expansion for higher-order correlation functions. This has allowed, while still retaining a semiclassical approximation for the pumped mode, to consider the full interacting quantum dynamics of the sidebands. The essential advantage of QuantumCumulants, with respect to other methods of quantum dynamics simulation such as QuTiP, is that it does not require any a priori cutoff on the dimension of the Hilbert space. This makes it particularly well-suited to exploring regimes where the expected occupations are very large but still finite, so that finite-size effects can still be relevant. Our simulations at second- and fourth-order cumulant expansion confirm we are operating in the described large-occupation limit. The simulation parameters are reported in Tab.~\ref{tab:simparameters}. We have then explored a range of detunings $\delta_0$ approaching the expected threshold value from higher values, starting from the stability region and then entering the bistability one.

\begin{table}
 \centering
 \begin{tabular}{|c|c|c|c|c|c|}\hline
 $\gamma$ & $\zeta_2/\gamma$ & $\zeta_2 l_\text{exp}^2/(2\gamma)$ & $g/\gamma$ & $P_\text{in}$ & $\beta/\gamma$ \\\hline
  $2\pi\,109.8(6)\,\text{MHz}$ & 0.0020(2) & 0.15(2) & $2.8(3)\times10^{-9}$ & $24(2)\,\text{mW}$ & $1.2(1)\times 10^4$\\\hline
 \end{tabular}
 \caption{Parameters employed in the stability analysis and the simulations, and their uncertainties from measurement or manufacturer specification.}
 \label{tab:simparameters}
\end{table}

\begin{figure}[h!]
\includegraphics[width=1\linewidth, keepaspectratio]{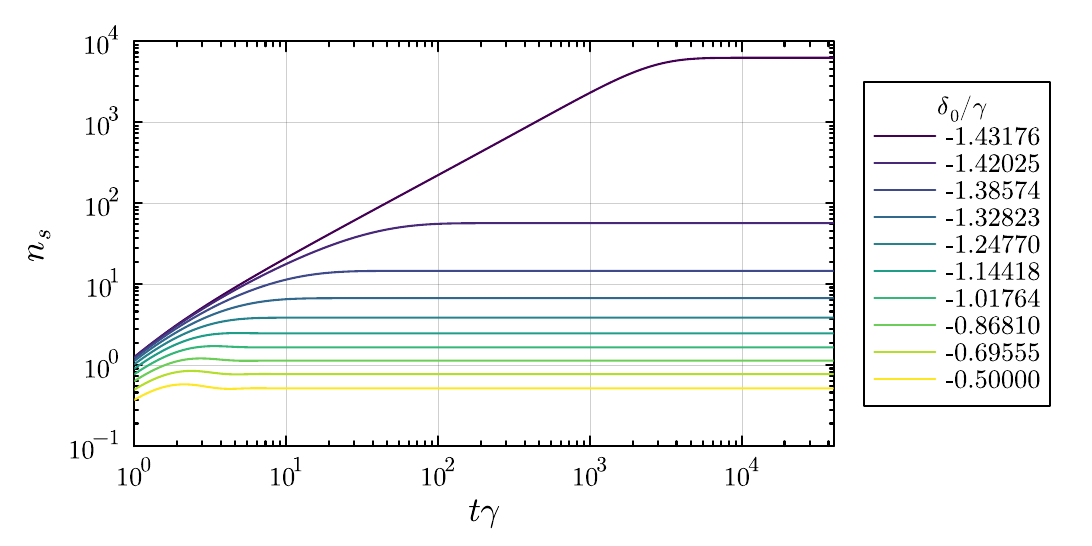}
\caption{Simulated evolution of the occupation $n_s(t)$ of the signal mode for input power $P_{\text{in}}=24\,\text{mW}$ for various cold-cavity detunings close to the OPO threshold.\label{fig:ns}}
\end{figure}

Using the library function \texttt{meanfield}, we have calculated the equations of motion for the quadratic operators appearing in \eqref{eq:su11generatorsEoM} at second-order cumulant truncation. The function \texttt{complete} is then instrumental in including equations of motion for all other operators needed to get a closed system of differential equations. We then have used a standard numerical solver to obtain time evolution starting from null initial conditions. In Fig.~\ref{fig:ns} we show the simulated evolution of the signal occupation $n_s(t)=\langle{\hat{a}_s^\dagger(t)\hat{a}_s(t)\rangle}$. The steady value $n_s$ increases with decreasing cold-cavity detuning $\delta_0$ with a faster rate as we get near the threshold detuning. The transient time scales as $1/\gamma$ in the regime where $\bar{\rho}$ is imaginary, whereas it starts being dominated by the factor $1/(\gamma -2\bar{\rho})$ in the regime of real $\bar{\rho}$ as we get nearer to the threshold. In order to approach the steady state value we need then to evolve our system over a time $\sim 1/(\gamma -2\bar{\rho})$.

\begin{figure}[h!]
\includegraphics[width=1\linewidth, keepaspectratio]{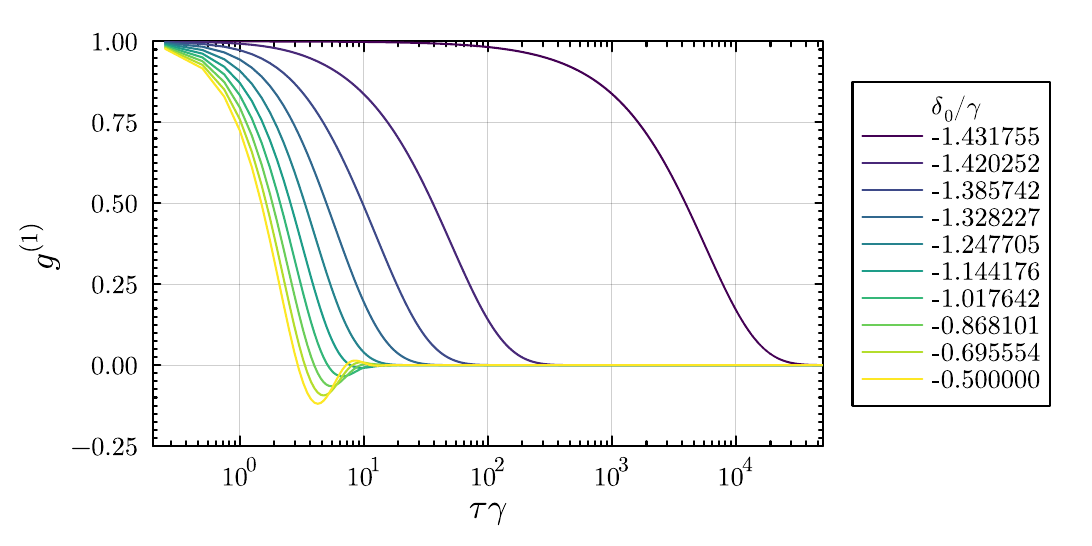}
\caption{Simulated first-order correlation function of the signal mode in the stationary regime $g^{(1)}(\tau)$ for input power $P_{\text{in}}=24\,\text{mW}$ for various cold-cavity detunings close to the OPO threshold.\label{fig:g1}}
\end{figure}

\begin{figure}[h!]
\includegraphics[width=1\linewidth, keepaspectratio]{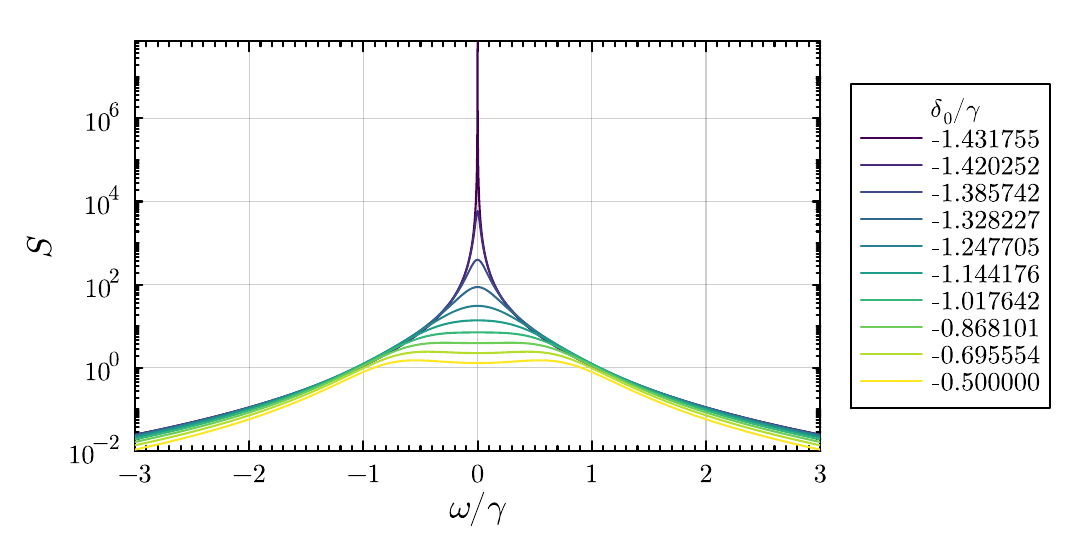}
\caption{Simulated spectrum of the signal mode in the stationary regime $S(\omega)$ for input power $P_{\text{in}}=24\,\text{mW}$ for various cold-cavity detunings close to the OPO threshold.\label{fig:S1}}
\end{figure}

The function \texttt{CorrelationFunction} is then used to compute equations of motion at second order for the first-order correlation functions, under the assumption the system is initialized in the steady state. Initial conditions are therefore given by the steady-state values for the quadratic operators. Results are shown in Fig.~\ref{fig:g1}. We can clearly observe the transition from the imaginary $\bar{\rho}$ regime, characterized by oscillations which are damped over a time $\sim 2/\gamma$, to the real $\bar{\rho}$ regime, where exponential decay occurs over a time $\sim 1/(\gamma/2 -\bar{\rho})$ in the proximity of the threshold. The corresponding spectrum, as shown in Fig.~\ref{fig:S1}, has the form of the product of two Lorentzians, first centered on $\pm|\bar{\rho}|$ and with equal FWHM $\gamma$, and then both centered on $\omega=0$, but with widths $\gamma\pm 2\bar{\rho}$. By evolving $G^{(1)}$ over a sufficiently long time, we are able to achieve a frequency resolution fine enough to observe the linewidth narrowing $\sim \gamma -2\bar{\rho}$ as we approach the threshold.

\section{Fit procedure\label{SMSec:fitg1}}

The theory curve and shaded region in Fig.~4 in main text is obtained by fitting the simulated data shown in Fig.~\ref{fig:g1}. The expression for the unnormalized first-order correlation function is shown in Eq.~\eqref{Eq:g1_general}, with $\bar{\rho} = \sqrt{g^{2}n_0^2 - (\delta_s + 2gn_0)^2}$ as defined in the main text. As analyzed, $n_0 = n_0(\delta_s)$ so that scanning the detuning parameter is possible to continuously tune the parametric gain $\bar{\rho}$ from purely imaginary to zero to purely real. We use these three limits to describe the different regimes of the system. 

For imaginary $\bar{\rho} = i\Omega$, then $\cosh \rightarrow \cos $ and $\sinh \rightarrow i\sin$. Using $a\cos(x) + b\sin(x) = C \cos(x + \phi)$, with $C = \sqrt{a^2 + b^2}$ and $\phi = -\arctan(b/a)$ we get:

\begin{align}
 f_1 = \frac{g^{2}n_0^2}{2\Omega\sqrt{(\gamma/2)^2 + \Omega^2}} e^{-\tau \gamma/2} \cos\left(\Omega\tau - \arctan\left(\frac{\gamma}{2\Omega}\right)\right).
\end{align}

This function describes a single exponential decaying with a characteristic time $2/\gamma$, with an additional modulation on top. Hence, we expect $\gamma_i \rightarrow\gamma$. 

For real $\bar{\rho}$ we can re-write Eq.~\eqref{Eq:g1_general} as:
\begin{align}
 f_2 = \frac{g^{2}n_0^2}{2\bar{\rho}} \left( \frac{e^{-\tau (\gamma - 2\bar{\rho})/2}}{\gamma - 2\bar{\rho}} - \frac{e^{-\tau (\gamma + 2\bar{\rho})/2}}{\gamma + 2\bar{\rho}}\right).
\end{align}

For decreasing $\delta_s$ (increasing $n_0$) the $\gamma/2 - \bar{\rho}$ contribution dominates, so that the function tends to the limit of a single exponential. For this reason we use $\gamma_i \sim \gamma - 2\bar{\rho}$, where $\gamma_i$ is the width of a single sideband mode (e.g. idler) in the approximation of a single Lorentzian spectrum. 
A particular case of real $\bar{\rho}$ is $\bar{\rho} \rightarrow 0$. Here the function becomes:

\begin{align}
 f_2(\bar{\rho} = 0) = \frac{2g^{2}n_0^2}{\gamma^2} e^{-\tau (\gamma/2)} \big( 1 + \tau\gamma/2\big).
\end{align}

This transition represents a bifurcation point where the system’s solution undergoes a qualitative change. Because the parameter $\bar{\rho}$ is defined by a square root, the transition at $\bar{\rho} = 0$ is characterized by a cusp of infinite slope in the derivative.
This is a bifurcation point where the fitting function changes from $f_1$ to $f_2$ (the slope is given by the derivative of a square root function). For this reason, for each data point we first discriminate the regime and then apply the corresponding functional form accordingly. 

The discrimination is performed by evaluating the $\chi$-square parameter associated with a single exponential fit:

\begin{align}
f_{\text{fit}} =
\begin{cases}
f_1 \sim A_0 e^{-\gamma_i \tau / 2} \cos(\Omega\tau + \phi) & \text{with }\gamma_i \rightarrow \gamma, \text{ if } \chi > \bar{\chi} \\
f_2 \sim A_0 \left( \frac{e^{-\gamma_i \tau / 2}}{\gamma_i} - \frac{e^{-\gamma_i^+ \tau / 2}}{\gamma_i^+} \right) & \text{with } \gamma_i \sim \gamma - 2\bar{\rho}, \text{ if } \chi < \bar{\chi} 
\end{cases}
\label{Eq:fit_function_g1_SM}
\end{align}
where $\bar{\chi}$ is a manually chosen critical value, $A_0$ is an amplitude parameter and we have introduced $\gamma_i^+ = \gamma + 2\bar{\rho}$. 
A similar procedure is carried out for experimental data. In particular, for the normalized second-order correlation function we still assume the equivalence $g^{(2)}(\tau) = 1 + |g^{(1)}(\tau)|^2$, with $g^{(1)}(\tau)$ given by Eq.~\eqref{Eq:fit_function_g1_SM}. 
While even in the presence of experimental noise, the difference between a single exponential decay and a modulated (oscillatory) signal is well-defined, the overdamped (real $\bar{\rho}$) regime presents a greater challenge. In fact, the subtle difference between a pure mono-exponential decay and a bi-exponential composition, comprising a dominant slow component and a lower-amplitude fast component, often falls below the resolution limit of the signal-to-noise ratio.

\begin{figure}[h!]
\includegraphics[width=1\linewidth]{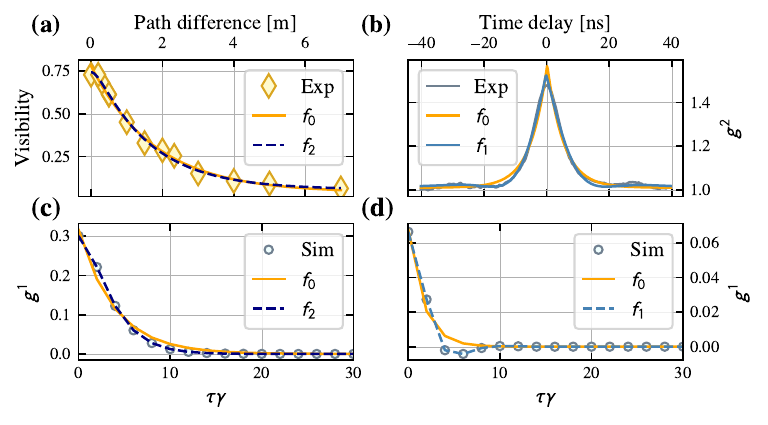}
\caption{Examples of fits to extrapolate the quantity $\gamma_i$. The orange solid line shows the fit performed assuming a single exponential decay for $G^{(1)}$. The dashed light-blue line uses the oscillatory $G^{(1)}(\tau)$ function ($f_1$), while the dashed dark-blue uses $g^{(1)}(\tau) = f_2$. 
a) Interferometry data for $\delta_\text{eff} = 1.31 \gamma$. $\gamma_i(f_0) = 54(5)\,\text{MHz}$, while $\gamma_i(f_2) = 62(5)\,\text{MHz}$.
b) $g^{(2)}(\tau)$ measurement for $\delta_\text{eff} = 1.39 \gamma$. $\gamma_i(f_0) = 117(2)\,\text{MHz}$, while $\gamma_i(f_1) = 72(2)\,\text{MHz}$. 
c) $g^{(1)}(\tau)$ simulation for $\delta_\text{eff} \sim 1.3 \gamma$. $\gamma_i(f_0) \sim 55\,\text{MHz}$, while $\gamma_i(f_2) \sim 104\,\text{MHz}$.
d) $G^{(1)}(\tau)$ simulation for $\delta_\text{eff} \sim 1.39 \gamma$. $\gamma_i(f_0) \sim 131\,\text{MHz}$, while $\gamma_i(f_1) \sim 110\,\text{MHz}$.
\label{fig:fitfunc} }
\end{figure}

The following showcases the two distinct physical regimes separated by the bifurcation point ($\bar{\rho} = 0$), which corresponds to an effective detuning of $\delta_\text{eff} \approx 1.31 \gamma$. In Fig.~\ref{fig:fitfunc}(a-c), we present experimental and simulated data for small detunings near this critical point, where the $g^{(1)}(\tau)$ function is described by the bi-exponential form $f_2$. Proximity to the bifurcation ensures that the two decay components have comparable weights, providing the ideal condition to distinguish a bi-exponential transition from a simple mono-exponential decay. While this distinction is clearly visible in simulation (panel c), experimental noise renders the two models compatible (panel a). In the frequency domain, this regime resolves into two centered Lorentzians: one narrow and high-amplitude, and one broad and significantly suppressed. As the system approaches the OPO threshold ($\bar{\rho} \to \gamma/2$), this spectral imbalance intensifies rapidly. Because the narrow component becomes overwhelmingly dominant, the error introduced by a mono-exponential approximation becomes negligible, making the selection of the complex fitting function less critical than in the oscillatory case.

In contrast, the oscillatory regime is characterized by the $f_1$ function, describing a double-peaked power spectrum where two distinct poles, each with a linewidth of $\gamma$, are separated by a spectral distance $2\Omega$. Approximating this modulated decay as a mono-exponential is mathematically equivalent to fitting this dual-peaked structure with a single, wide Lorentzian envelope. While the difference is most distinct in the simulated $g^{(1)}(\tau)$, we find that the experimental $g^{(2)}$ results yielded by the two models are also incompatible. Consequently, utilizing the correct piecewise fitting function is essential in this regime to ensure accurate parameter extraction and to avoid artificial spectral broadening, particularly for long acquisitions with high signal-to-noise ratios.

\section{Coupling constant $g$ vs temperature \label{SM:gvsT}}

The effective coupling coefficient $g$ between ring resonator modes is given by~\cite{Mat05}
\begin{equation}
 g = c\hbar \frac{n_2}{n^2} \frac{\omega_0^2}{V_\text{eff}}
\end{equation}
where $n_2$ is the Kerr coefficient; $n$ is the linear refractive index (phase index);
$\omega_0$ is the frequency of the pumped mode and $V_\text{eff}$ is the effective volume.
In particular, for the $\ell_0$-th mode $\omega_0 = 2\pi \times \ell_0 c/(n 2\pi r)$, with $r$ the radius of the ring. So $g$ becomes
\begin{equation}
 g = ch \frac{n_2}{n^4} \frac{\ell_0^2 c^2}{A_\text{eff} (2\pi r)^3}
\end{equation}
with $A_\text{eff}$ effective mode area. 

The temperature dependence of the coupling coefficient $g$ is governed by:
\begin{equation}
\frac{1}{g} \partial_T g \approx -4 \alpha_n - 3\alpha_r,
\end{equation}
where $\alpha_n = \frac{1}{n}\partial_T n$ is the normalized thermo-optic coefficient and $\alpha_r = \frac{1}{r}\partial_T r$ is the coefficient of thermal expansion. Here, we have neglected the temperature dependence of the effective mode area $A_\text{eff}$. For $\text{Si}_3\text{N}_4$ at telecom wavelengths, typical values are $\alpha_n \approx 2.45 \times 10^{-5} \, ^\circ\text{C}^{-1}$~\cite{Arb:13} and $\alpha_r \approx 3.27 \times 10^{-6} \, ^\circ\text{C}^{-1}$~\cite{Tien:12}. Since the thermo-optic term dominates the thermal variation of the parameter $g$, the total fractional change $\partial_T g / g$ remains on the order of $10^{-5} \, ^\circ\text{C}^{-1}$. 
We can verify the magnitude of this effect by estimating the internal temperature rise during the pump frequency sweep shown in Fig.~\ref{fig:det_eff_res}. By extrapolating the linear (cold-cavity) region of the detuning curve, we find that the detuning at the final pump frequency would be approximately \qty{-560}{MHz} in the absence of thermal effects. Since the measured effective detuning reaches \qty{80}{MHz}, we conclude that the resonance frequency shifted by $\sim\qty{640}{MHz}$ due to absorbed pump power. Given the dominance of the thermo-optic effect, this spectral shift corresponds to a temperature variation of approximately $0.14\,^\circ\text{C}$. Such a minor increase results in a fractional change in $g$ of only $\sim 0.001\%$, justifying our treatment of the nonlinear coupling strength as a constant parameter throughout the transition.

\end{document}